\documentclass[a4paper,fleqn,usenatbib,useAMS]{mnras}

\usepackage{graphicx}	
\usepackage{amsmath}	
\usepackage{amssymb}	
\usepackage{multicol}        
\usepackage{bm}		
\usepackage{pdflscape}	
\usepackage{float,textcomp}
\usepackage{natbib}
\usepackage{enumerate}
\usepackage{lscape}
\usepackage{afterpage}
\usepackage{pdflscape}
\usepackage{subfigure}
\usepackage{color}
\usepackage[english]{babel}
\usepackage[T1]{fontenc}

\title[The Void Galaxy Survey: Star Formation Properties]{The Void Galaxy Survey: Star Formation Properties}

\author[B. Beygu et al.]{B. Beygu,$^{1,6}$\thanks{E-mail: burcu.beygu$@$nwu.ac.za}
K. Kreckel,$^{2}$
J. M. van der Hulst,$^{1}$
T. H Jarrett,$^{3}$
R. Peletier,$^{1}$ 
\newauthor
R. van de Weygaert,$^{1}$
J. H. van Gorkom,$^{4}$
M. A. Aragon-Calvo$^{5}$
\\
$^{1}$Kapteyn Astronomical Institute, University of Groningen, PO Box 800, 9700 AV Groningen, the Netherlands\\
$^{2}$Max Planck Institute for Astronomy, K\"{o}nigstuhl 17, 69117 Heidelberg, Germany\\
$^{3}$Astronomy Department, University of Cape Town, Rondebosch 7700, Cape Town, South Africa\\
$^{4}$Department of Astronomy, Columbia University, Mail Code 5246, 550 West 120th Street, New York, NY 10027, USA\\
$^{5}$University of California, Riverside, USA\\
$^{6}$Physics and Centre for Space Research, North-West University, Potchefstroom, South Africa
}

\pubyear{2016}

\begin{document}
\label{firstpage}
\pagerange{\pageref{firstpage}--\pageref{lastpage}}
\maketitle

\begin{abstract}

We study the star formation properties of 59  void galaxies as part of the Void Galaxy Survey (VGS). Current star formation rates are derived from $\rm{H\alpha}$ and
recent star formation rates from  near-UV imaging. In addition, infrared 3.4 $\rm{\mu m}$, 4.6 $\rm{\mu m}$, 12 $\rm{\mu m}$ and 22 $\rm{\mu m}$ WISE emission is 
used as star formation and mass indicator. Infrared and optical colours show that the VGS sample displays a wide range of dust and metallicity properties. We 
combine these measurements with stellar and $\rm{H\textsc{i}}$ masses to measure the specific SFRs ($\rm{SFR/M_{*}}$) and star formation efficiencies 
($\rm{SFR/M_{H\textsc{i}}}$). We compare the star formation properties of our sample with galaxies in the more moderate density regions of the cosmic web, 
'the field'. We find that specific SFRs of the VGS galaxies as a function of stellar and $\rm{H\textsc{i}}$ mass are similar to those of the galaxies in these 
field regions. Their $\rm{SFR\alpha}$ is slightly elevated than the galaxies in the field for a given total $\rm{H\textsc{i}}$ mass. In the global star formation
picture presented by Kennicutt-Schmidt, VGS galaxies fall into the regime of low average star formation and correspondingly low 
$\rm{H\textsc{i}}$ surface density. Their mean $\rm{SFR\alpha/M_{HI}}$ and $\rm{SFR\alpha/M_{*}}$ are of the order of $\rm{10^{-9.9}}$ $\rm{yr^{-1}}$. We conclude 
that while the large scale underdense environment must play some role in galaxy formation and growth through accretion, we find that even with respect to other 
galaxies in the more mildly underdense regions, the increase in star formation rate is only marginal.
\end{abstract}




\newpage

\section{Introduction}
\label{section:intro}

Voids are prominent features of the cosmic web \citep{weyplaten2011}. Formed from primordial
underdensities they now occupy a major fraction of the volume of the universe, surrounded
by denser filaments, walls and sheets. They are the most underdense regions in the universe and 
are the most pristine environments where galaxy evolution will have progressed slowly, without 
the dominant and complex influence of the environment. Voids therefore are extremely well suited  
for assessing the role of the environment in galaxy evolution, as here the galaxies are expected 
not to be affected by the complex processes that modify galaxies in high density environments. The void environment covers the lowest density 
environments found in the universe, though some voids do approach similar
(and still low) densities as found in tenuous filaments and walls \citep{marius2014}. These authors investigate the dark matter halo 
distribution in the various cosmic web components and demonstrate that in
voids very few dark matter halos more massive than $10^{11} \rm{ M_{\odot}}$
exist, confirming the idea that high stellar mass objects are rarely
expected in these environments. 

In order to get a good picture of galaxies in voids the Void Galaxy Survey (VGS) was designed, a
multi-wavelength study of $\sim$ 60 galaxies in geometrically defined voids
\citep{stanonik2009,rien2011,kreckel2011,kreckel2012}. Previous papers have focused on the $\rm{H\textsc{i}}$
properties of galaxies in voids and found that the voids contain a population of galaxies that are 
relatively $\rm{H\textsc{i}}$ rich of which many present evidence for ongoing gas accretion,
interactions with small companions and filamentary alignments \citep{kreckel2011, kreckel2012,
beygu2013}. Void galaxies in general have small stellar masses ($\le 3 \times 10^{10}$ $\rm{ M_{\odot}}$). This is consistent with previous studies analysing the optical properties of 
void galaxies. These   
show that void galaxies are in general small, star forming blue 
galaxies and have late morphological types
\citep{szomoru1996,kuhn1997,popescu1997,karachentsev1999,grogin1999,grogin2000,
rojas2004,rojas2005,tikhonov2006,patiri2006a,patiri2006b,ceccar2006,wegner2008,kreckel2011,kreckel2012,moorman2014,moorman2015}.
Conclusions on the role of void environment relative to the field, however, are not clear. \cite{patiri2006b}, using the SDSS DR4 data, show that void galaxies have 
the same specific
star formation rates at a fixed colour as their comparison sample of field galaxies. Similarly \cite{penny2015} examined the properties of void galaxies in the Galaxy and Mass 
Assembly (GAMA, \cite{driver2011}) survey. They found that void galaxies with stellar mass $\rm{M_{*} >}$ $5 \times 10^{9} M_{\odot}$ have ceased forming stars and their infrared
colour distribution show a wide range of star formation activity. \cite{rojas2004,rojas2005}, using equivalent widths of $\rm{H\alpha}$, 
[O\textsc{ii}], [N\textsc{ii}], 
$\rm{H_{\beta}}$ and [O\textsc{iii}] from SDSS DR4, argued that void galaxies have similar star formation rates as field 
galaxies\footnote{In their method field galaxies are indicated by the name "wall galaxies" as defined in \cite{hoyle2002} }  
but since they have smaller stellar masses, their specific star formation rates are expected to be higher than field galaxies. \cite{ricci2014} 
reported that void galaxies appear to form stars more efficiently then galaxies living in the void shells and the general galaxy population based on their
sample form the SDSS DR7. They also find that the star formation rate is insensitive to the environment for their sample when only the star-forming galaxies are 
considered. \cite{moorman2016}, however, found contradicting results. For a subset sample of optically selected galaxies from the SDSS DR8 with $\rm{H\textsc{i}}$ detection they 
found that the specific star formation rate did not vary systematically with large scale environment. They also did not find any environmental dependence for the star formation rate per 
unit $\rm{H\textsc{i}}$ mass. They discuss in detail that the difference between the results of \cite{ricci2014} and theirs arises from the different definition of a void.
On the other hand, \cite{grogin2000} investigated the $\rm{H\alpha}$ equivalent 
widths of void galaxies selected from the 2dF. They found indications that void galaxies with companions to have more elevated star formation rates than comparable 
field galaxies, while the star 
formation rate of void galaxies without nearby companions varies little over the entire density range. A study of nearby cosmic voids within a distance of 40 Mpc, 
by \cite{el2013},
 found 48 late-type dwarf galaxies, with a median star formation rate per luminosity of 
$\rm{\sim 10^{-10}}$ $\rm{M_{\odot} yr^{-1} L^{-1}_{\odot}}$. Furthermore, physical properties of 26 
emission line galaxies in the Bootes Void have been investigated by \cite{cruzen2002}. They extended the sample used in the $\rm{H\alpha}$ 
imaging of the 12 Bootes Void galaxies investigated by 
\cite{peim1992}, and spectra of the 10 Bootes void galaxies studied by \cite{weis1995}. They did not include, however, additional Bootes Void galaxies reported by 
\cite{szo1996}, 
into their sample. By using equivalent widths of $\rm{H\alpha}$+[N\textsc{ii}], and line ratios between [O\textsc{iii}], $\rm{H\beta}$ and [S\textsc{ii}], 
\cite{cruzen2002} identified two extreme 
starbursts and 13 galaxies with elevated rates of star formation out of 26 galaxies.

In most of the previous studies \citep{rojas2004,rojas2005,hoyle2012,ricci2014,moorman2016}, 
star formation rate measurements were mostly based on aperture corrected spectra of large survey data without any information
about the star formation distribution throughout the galaxy. They are missing the combination of $\rm{H\alpha}$, UV, infrared and 21-cm neutral hydrogen imaging data 
through which star 
formation efficiencies, current and recent star formation properties of void galaxies can be directly measured and emission morphologies can be clearly investigated. 
Although, studies of 
\cite{peim1992,weis1995,szo1996} involve individual observations, their sample either suffer from selection effects (since the sample consist 
of mostly IRAS selected galaxies) or lack of completeness.  

In this study, we present, for the first time, star formation properties of void galaxies measured from systematic $\rm{H\alpha}$, near-UV, infrared and 21-cm neutral 
hydrogen imaging surveys. We assume $\rm{H_{0}=70 km s^{-1} Mpc^{-1}}$.

This study is part of the Void Galaxy Survey (VGS) \citep{kreckel2011,rien2011} and targets the galaxies selected for that study (59 targeted galaxies and 18 companions 
confirmed by either 21-cm neutral hydrogen or optical observations\footnote{The original sample was meant to contain 60 
galaxies. However, one of the galaxies, VGS\_28, has been removed from the sample since redshift information from the latest SDSS release was inconsistent with the 
previous SDSS data releases and placed it well outside void. It was also not detected in HI as redshift is outside the observing band. This reduces the number of targeted 
galaxies to 
59.}).
\section{The VGS sample: observations and reduction}
\label{sec:vgs}

\begin{figure*}
  \centering
\subfigure[]
{
  \includegraphics[scale = 0.55,angle=360]{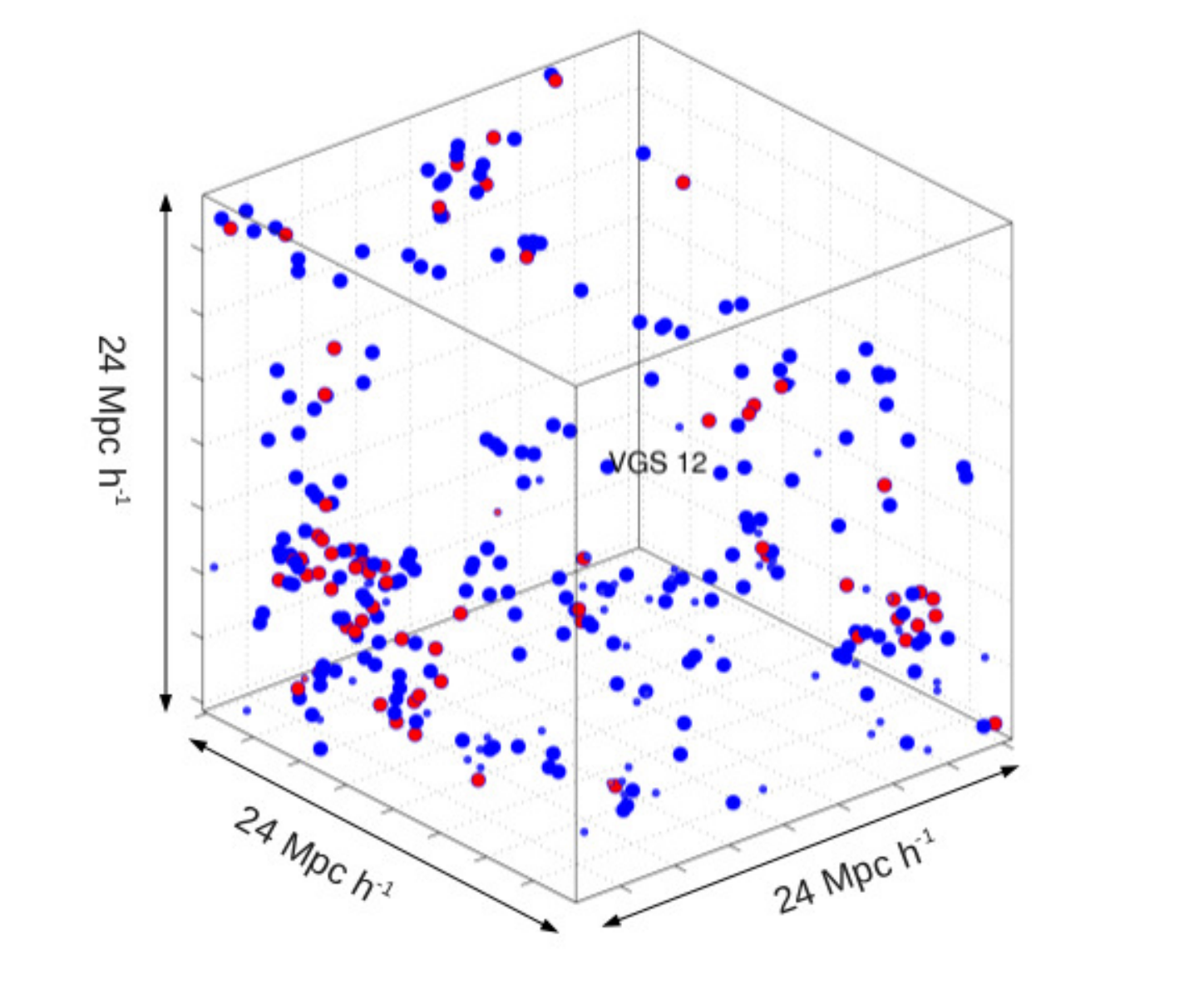}
\label{fig:first_sub}
}
\subfigure[]
{
  \includegraphics[scale = 0.34,angle=360]{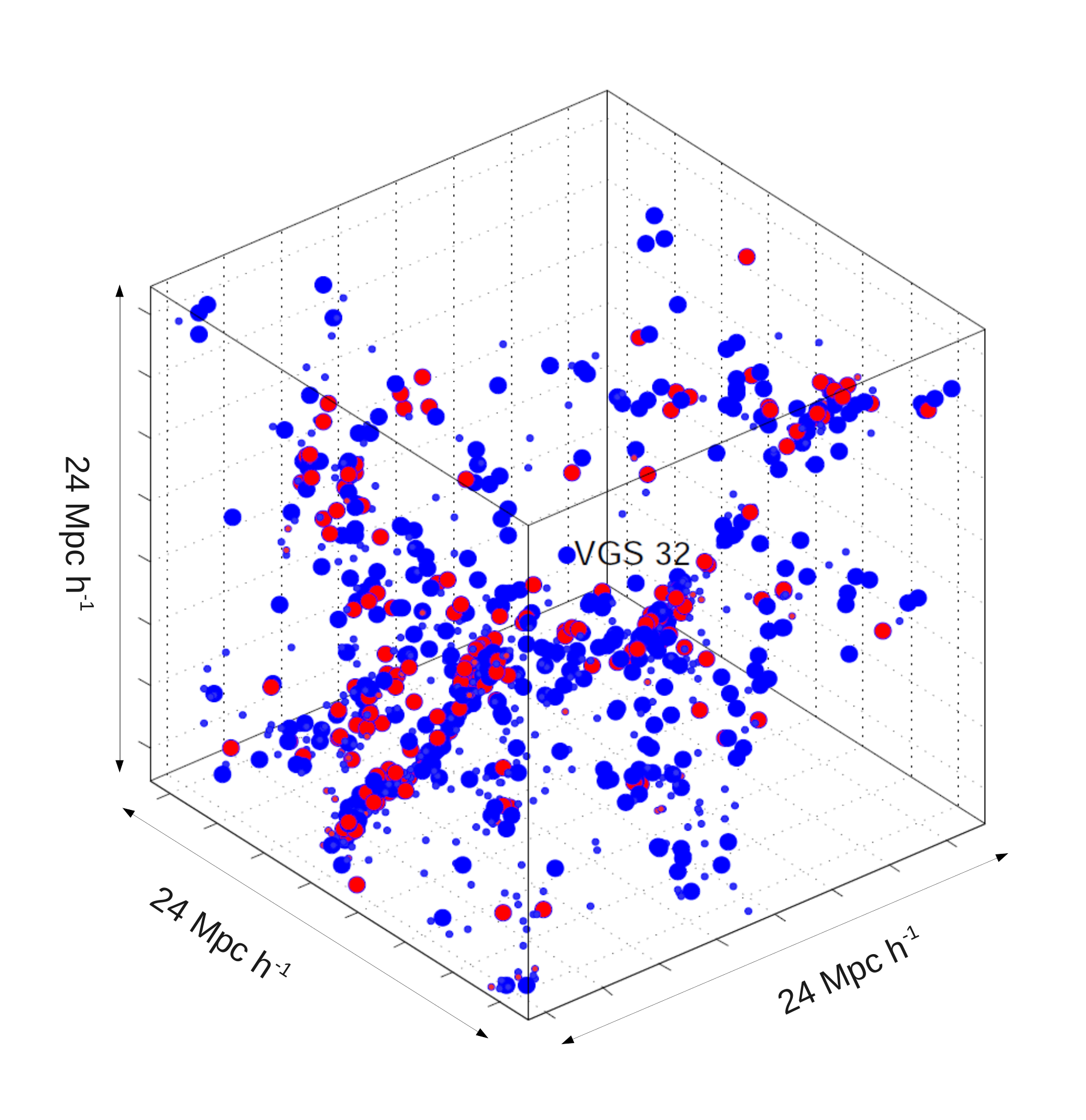}
\label{fig:second_sub}
}
\caption{The large-scale structure distribution of galaxies within 24 Mpc $\rm{h^{-1}}$ in comoving coordinates around VGS\_12 (a) and VGS\_32 (b). Surrounding galaxies are colour 
coded by $g-r$ colour, to be red if $g-r >$ 0.6 and blue if $g-r \leq$ 0.6. The symbol size indicates luminosity, with larger symbols if $M_{r}$ $<$ -18 and smaller 
symbols if $M_{r}$ $\geq$-18. $M_{r}$ is derived from the SDSS DR7 apparent $r$ model magnitudes. } 
  \label{figure:void_size}
\end{figure*}

Galaxies in the VGS have been selected from the Sloan Digital Sky Survey Data Release 7 (SDSS DR7) using purely geometric and topological  
techniques. The sample was selected on the basis of galaxy density maps produced by the Delaunay Tessellation Field Estimator 
(DTFE, \cite{schaap2000, weyschaap2009}) and the subsequent application of the Watershed Void Finder (WVF, \cite{platen2007}). The 
combination of DTFE maps 
with WVF detected voids allow us to identify void galaxies from the deepest interior regions of identified voids in the SDSS 
redshift survey. Our sample has a density contrast, $\rm{\delta \equiv \rho_{void}/\rho_{u}}$ -1, of less than -0.5, where $\rm{\rho_{u}}$ is the mean density.
In Figure~\ref{figure:void_size}, we present two examples displaying two different environments surrounding the voids in 
which two of the VGS galaxies reside. One
(Figure~\ref{figure:void_size}a) is the environment of the polar disk galaxy 
(VGS\_12) mentioned above, and the other (Figure~\ref{figure:void_size}b) is 
the environment of a flocculent galaxy (VGS\_32) sitting in a
void that is surrounded by a massive filament.

Our geometrically selected sample consists of small galaxies, with stellar mass less than $3 \times 10^{10}$ $\rm{ M_{\odot}}$. The $\rm{H\textsc{i}}$ 
mass range of our sample is between $\rm{10^7}$ to $\rm{10^{10}}$ $\rm{ M_{\odot}}$ and there redshifts range from 0.02 to 0.03. Most of them are blue star forming disk galaxies and many 
of them have companions and extended $\rm{H\textsc{i}}$ disks, which are often morphologically and kinematically disturbed \citep{kreckel2011,kreckel2012}. Two objects have been
 investigated in great detail. One of them is a polar disk galaxy (VGS\_12), 
with an $\rm{H\textsc{i}}$ disk, about nine times the extent of the stellar disk, rotating with an angle perpendicular to the stellar disk
 \citep{stanonik2009}. The second one, VGS\_31, is a system of three interacting galaxies forming a filamentary structure inside a void \citep{beygu2013}. 
 
The VGS galaxies have been observed in the B band, $\rm{H\alpha}$,  UV, near IR and in the
$\rm{H\textsc{i}}$. Narrow band $\rm{H\alpha}$ imaging
has been obtained using the Hiltner Telescope at the Michigan-Dartmouth-MIT Observatory (MDM). 
B band imaging has been gathered with the Isaac Newton 
Telescope (INT) at La Palma using the Wide Field Camera (WFC) Near UV (NUV) images have been 
taken with GALEX. Near-infrared images have been obtained from the WISE survey using Jarrett's pipeline
\citep{jarrett2013} to extract calibrated in the 3.4 $\rm{\mu}m$, 4.6 $\rm{\mu}m$, 12 $\rm{\mu}m$
and 22 $\rm{\mu}m$ bands. A detailed description of these data sets is given below. In an accompanying paper morphology and colour properties of the VGS galaxies
derived from these data will be discussed (Beygu et al. to be submitted).

\subsection{$\rm{H\small\textsc{I}}$ imaging}
 
The VGS galaxies have been observed with the Westerbork Synthesis Radio Telescope (WSRT) \citep{kreckel2012} in the maxi-short configuration providing
an angular resolution of $\rm{19^{\prime\prime} \times 32^{\prime\prime}}$. We observed 512 channels within a total bandwidth of 10 MHz, 
giving a Hanning smoothed velocity resolution of 8.6 km s$\rm{^{-1}}$. The 36$\rm{^\prime}$ full width half maximum of the WSRT primary beam and the given total bandwidth allowed us to probe 
a total volume covering $\sim$ 1.2 Mpc and 1200 $\rm{km}$ $\rm{s^{-1}}$ at 85 Mpc. Images for this paper were made with natural weighting to maximize 
sensitivity and cleaned to a level of 0.5 mJy beam$\rm{^{-1}}$ ($\rm{\sim}$1 $\rm{\sigma}$), reaching column density sensitivities of 2 $\rm{\times}$ 10$^{19}$
 cm$^{-2}$. $\rm{HI} $ detections and masses are given in Table 2 in \cite{kreckel2012} for the full sample of VGS galaxies.

\subsection{$\rm{H\alpha}$ imaging}

\begin{figure}
 \centering\includegraphics[scale=0.37]{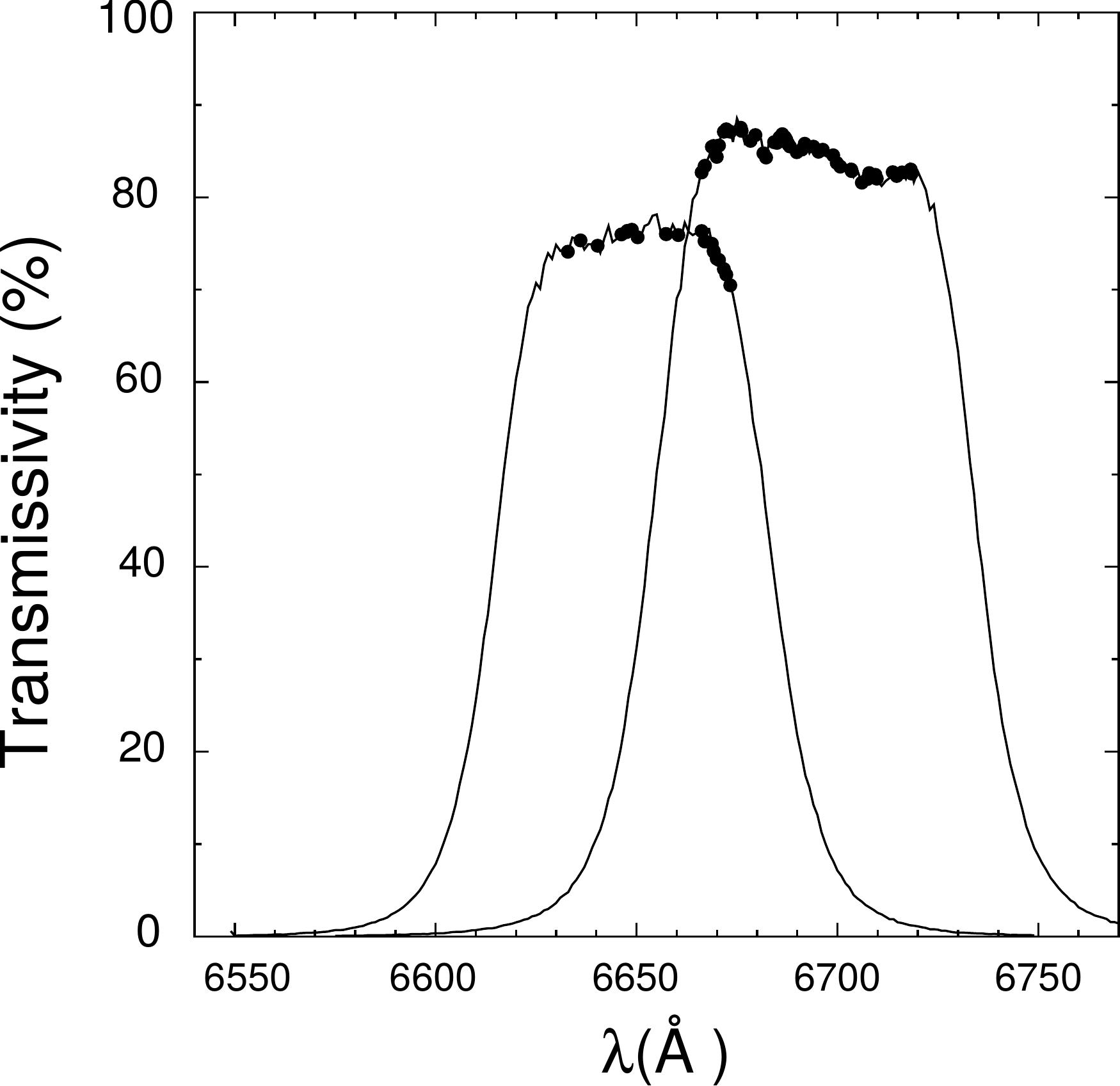} 
  \caption{The transmissivity of the  $\rm{H\alpha}$ filters with central wavelengths of 6649 $\rm{\AA}$
and 6693 $\rm{\AA}$. Filled circles mark the transmissivity for $\rm{H\alpha}$ at 
the redshift of the target galaxies.}
  \label{fig:3.2}
\end{figure}

$\rm{H\alpha}$ imaging has been obtained with the Echelle CCD in direct mode at the 2.4 m Hiltner Telescope. The redshifted $\rm{H\alpha}$ filters 
centred at 6649 $\rm{\AA}$ and 6693 $\rm{\AA}$ were used (Figure~\ref{fig:3.2}). To have a measure of the continuum, R (Harris) band imaging was performed for each object. The
total integration time for $\rm{H\alpha}$ (1800 seconds) and for the continuum (360 seconds) was spread over 3 exposures for the purpose of dithering and for facilitating cosmic ray detection. 
Spectrophotometric calibration stars were chosen from either  \cite{massey1988} or \cite{oke1990}.
59 VGS galaxies and 8 out of 17 companions have been observed in $\rm{H\alpha}$. In total 62 galaxies have been detected.

The data have been reduced using the standard IRAF\footnote{http://iraf.noao.edu/} procedures for CCD imaging. 
All the optical images were trimmed and overscanned followed by bias subtraction and flat fielding. All images from each filter were aligned 
and median combined. Each combined $\rm{H\alpha}$ and R band image was normalized by the integration time. The mean was calculated for an empty region in each image and the ratio of 
these means
was taken as scaling factor for scaling the continuum image before subtraction from the $\rm{H\alpha}$ image. The photometric calibration of the final $\rm{H\alpha}$ images 
was performed following the steps described in \cite{gavazzi2006} and references therein. Corrections for the atmospheric extinction and the airmass have been performed
in the standard way, where each spectrophotometric calibration star observation has been fitted using airmass and instrumental magnitudes to get the atmospheric extinction
coefficient. The final $\rm{H\alpha}$ images have been created by:

\begin{equation}
C_{ON} - nC_{OFF} = C_{NET},
\end{equation}
where n is the scaling factor or normalisation coefficient, $C_{ON}$ and $C_{OFF}$ are the flux in counts per second for the $\rm{H\alpha}$ and continuum
filters respectively and $C_{NET}$ is the net flux in counts after subtracting the continuum. After checking the mean of flux for the stars in the field, $n$ was
improved if necessary. 

Flux calibration has been performed using spectrophotometric standard stars, following the basic photometry equation relating magnitude, $m$ and
flux, $F_{star}$:

\begin{equation}
 m = -2.5log(F_{star}) + C_{star},
\end{equation}
This yields:

\begin{equation}
-2.5log(F_{star}) = -2.5log(C_{star}) + m_{ZP} - \kappa sec(z),
\end{equation}
where $m_{ZP}$ is magnitude zero point, $\rm{\kappa}$ is the atmospheric extinction coefficient, $sec(z)$ is the airmass, $C_{star}$ is the standard star's total counts in an
aperture and $F_{star}$ is the flux density of the standard star calculated as:

\begin{equation}
 F_{star} = \int S(\lambda) R_{ON}(\lambda) \; \mathrm{d}\lambda,
\end{equation}
where $S(\lambda)$ is the spectral energy distribution of the standard star and $R_{ON}(\lambda)$ is the transmissivity of the $\rm{H\alpha}$ filter
as a function of wavelength. Then the total $\rm{H\alpha}$ flux, $F(H_{\alpha})_{0}$ ( contaminated by [N\textsc{ii}] ) for a galaxy is:

\begin{eqnarray}
 F(H_{\alpha})_{0} =10^{log(\frac{F_{star}}{C_{star}})- 0.4 \kappa sec(z)}  \times \dfrac{C_{NET}}{ R_{ON}(z)}
\end{eqnarray}   
where $R_{ON}(z) = R_{ON}(Filter \times (1+z))$ is the transmissivity of the $\rm{H\alpha}$ filter as a function of a galaxy's redshift. Correction for the 
contamination
of the $\rm{H\alpha}$ line emission in the continuum filter yields: 

\begin{equation}
F(H_{\alpha}) = F(H_{\alpha})_{0} \times \; \left[ 1+ \dfrac{\int R_{ON}(\lambda) \mathrm{d}\lambda}{\int R_{OFF}(\lambda) \mathrm{d}\lambda}\right], 
\end{equation}
where $F(H_{\alpha})$ is the line corrected flux and $R_{ON}(\lambda)$ and $R_{OFF}(\lambda)$ are the transmissivities of the $\rm{H\alpha}$ and the 
continuum filters respectively.

\begin{figure}
\centering\includegraphics[scale=0.35]{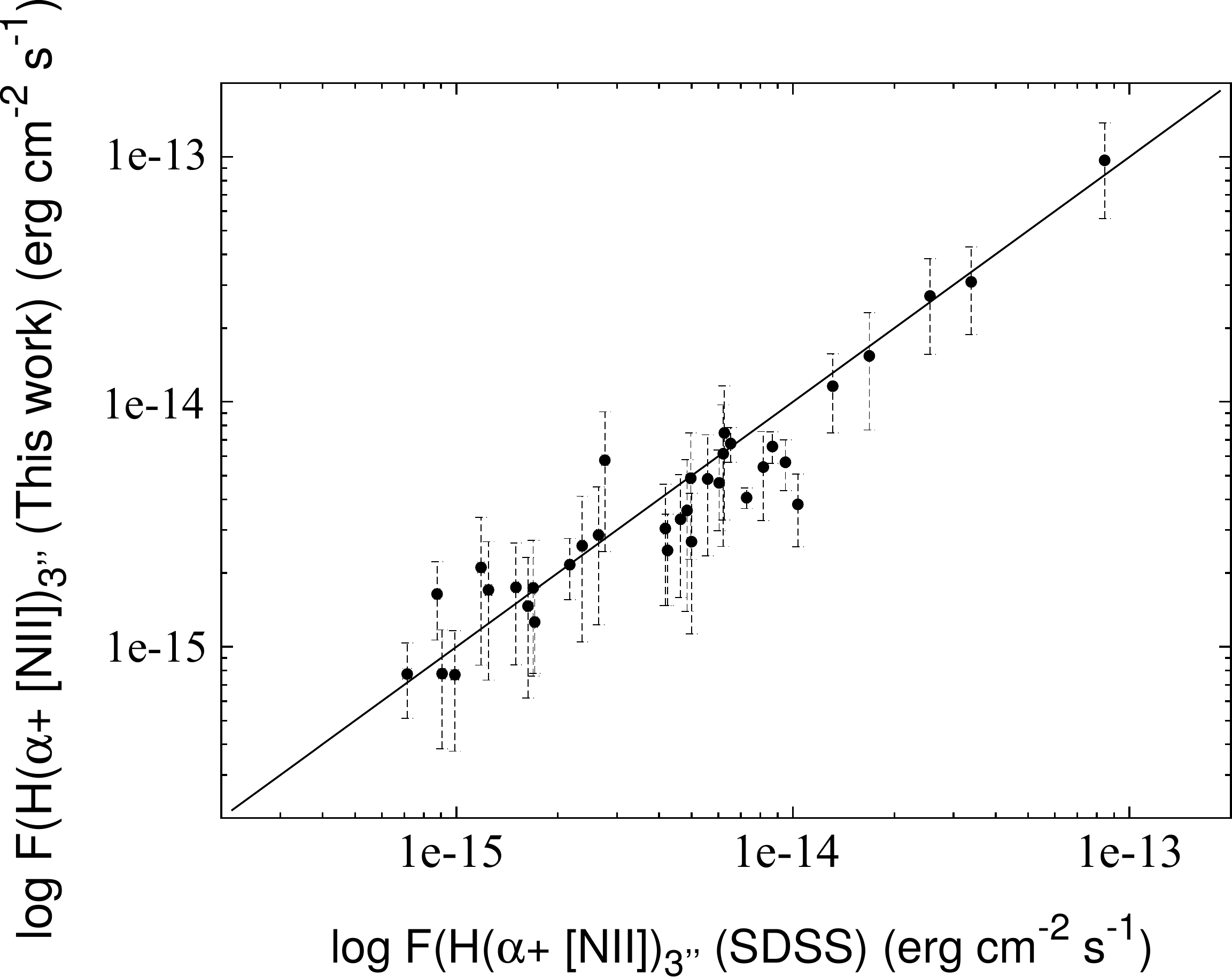} 
  \caption{Comparison between the $\rm{H\alpha}$ + [N\textsc{ii}] fluxes extracted in 3 arcsecond aperture from this work and the $\rm{H\alpha}$ + [N\textsc{ii}] lines
given by the MPA/JHU catalogue in 3 arcsecond fiber spectrum. The solid line is the identity line. Only the galaxies with high S/N $>$ 3 detection are plotted. Errors on SDSS
 $\rm{H\alpha}$ + [N\textsc{ii}] fluxes are significantly small compared to this work, therefore they are not visible in the plot. }
  \label{fig:3.1}
\end{figure}

The $\rm{H\alpha}$ line fluxes measured within the 3$''$ fiber apertures of the SDSS pipeline compare very well with the $\rm{H\alpha}$ 
fluxes extracted in  3$''$ apertures from the MDM  $\rm{H\alpha}$ narrow band images (Figure ~\ref{fig:3.1}), thus confirming the calibration. Both $\rm{H\alpha}$ flux measurements 
include a contribution from the [N\textsc{ii}] lines.

\subsubsection{Correction for [N\textsc{ii}] line contamination }

The contribution of the [N\textsc{ii}] line to the observed flux has been estimated using the expression from \cite{kennicutt2008}: 

\begin{equation*}
log ([N\textsc{ii}]/ H_{\alpha})=
\end{equation*}
\begin{equation*}
\begin{cases} (-0.173 \pm 0.007) M_{B} - (3.903 \pm -0.137) & \text{if $M_{B} > -21$,}
\\
 0.54 &\text{if $M_{B} \leq -21$.}
\end{cases}
\end{equation*}

and has been subtracted from the $\rm{H\alpha}$
flux. $\rm{M_{B}}$ has been calculated from the SDSS \textit{g} magnitudes 
following \cite{gavazzi2012}.

\subsection{Near UV imaging}

Near UV data have been obtained with The Galaxy Evolution Explorer (GALEX) satellite, 
primarily collected under the Guest Investigator program 061 in Cycle 6. 
Observations have exposure times of $\sim$ 1600 seconds.
The data was calibrated using the standard GALEX pipelines. 
NUV magnitudes ($\rm{m_{AB}}$) 
are taken from the GALEX pipeline catalogue as measured with a Kron aperture \citep{kron1980}.
The flux density $\rm{f_{\upsilon}}$ is given by:

\begin{eqnarray}
 f_{\upsilon}[erg\;s^{-1} cm^{-2} Hz^{-1}] = 10^{-0.4 (m_{AB}+48.6)}
\end{eqnarray}

\noindent
59 VGS galaxies and 11 companions have been detected in the near UV.

\subsection{Ancillary data: SDSS and WISE}

In addition to our own observations we also make use of archival data. Before summarising our observations and data analysis steps, we describe the archival data we used to complement
our analysis.

\subsubsection{SDSS photometric and spectroscopic data}
\label{sec:sdss}

$\rm{H\alpha}$, $\rm{H_{\beta}}$, [N\textsc{ii}] and [O\textsc{iii}] emission line fluxes, 4000-$\rm{\AA}$ break values ($\rm{D_{n}(4000)}$),($g-r$) colours and
the gas-phase metallicities (12+Log(O/H), following \cite{tremonti2004}) of the 
VGS galaxies have been taken 
from the MPA/JHU catalogue for SDSS DR7. These have been measured from the SDSS DR7 spectra which cover the central 3$''$ of each object. Average distance of our sample is $\sim$ 80 Mpc.
At this distance 3$''$ corresponds to a physical scale of 1.6 kpc. Absolute B magnitudes ($\rm{M_{B}}$) of the VGS 
galaxies have been calculated using the SDSS \textit{g} magnitudes, following \cite{gavazzi2012} and 
references therein.  We used  the
 $\rm{H\alpha}$ and  $\rm{H_{\beta}}$ line fluxes to measure Balmer decrements of VGS galaxies as described in
section 3.5.1.

We used [N\textsc{ii}] and [O\textsc{iii}] line fluxes in order to create a Baldwin, Phillips \& Terlevich (BPT) diagram 
(Figure~\ref{figure:9}). For this, only galaxy spectra whose measured line amplitudes are 5 times the size of the residual noise have been used. 
Among these galaxies, only one galaxy spectrum (VGS\_24), has been fitted separately using the pPXF \citep{cappel2004} and GANDALF \citep{sarzi2006} packages 
since it is classified as broadline object\footnote{In the SDSS spectroscopic catalogue if any galaxies or quasars have lines detected at the 10$\rm{\sigma}$ level with $\rm{\sigma}$ $>$
 200 km s$\rm{^{-1}}$ at the 5$\rm{\sigma}$ level, the indication "BROADLINE" is appended to their subclass.} and the SDSS pipeline has problems with fitting the spectra of 
 these objects. 
 
The MPA/JHU catalogue for SDSS DR7 \footnote{The MPA-JHU catalogue is publicly available and may be downloaded
at http://www.mpa-garching.mpg.de/SDSS/DR7/archive} provides stellar mass estimates that are obtained using fits to the \textit{ugriz} photometry following the methods 
described in \cite{kauffmann2003} and \cite{salim2007}. In \cite{kreckel2012} we have adopted the stellar masses given by the MPA/JHU catalogue for the full sample of the 
VGS galaxies. In this study we also use the WISE (the Wide-field Infrared Survey Explorer) [3.4] and [4.6] data to determine the stellar mass of the VGS galaxies in order to be consistent with the stellar mass measurements of 
the comparison sample for which stellar masses are determined via infrared data. A comparison between stellar masses determined by the WISE measurements
and the MPA/JHU catalogue is given in section~\ref{sec:wise_sm}.

\subsubsection{WISE data}

WISE 3.4 $\rm{\mu}m$, 4.6 $\rm{\mu}m$, 12 $\rm{\mu}m$ and 22 $\rm{\mu}m$ measurements have been retrieved 
from the WISE All-sky source catalogue \citep{wright2010}. Resolved sources (the WISE angular beam is 
about 6 arcsec) are measured using code and tools developed for the WISE photometry pipelines
\citep{jarrett2011,jarrett2013,cutri2012}. More details may be found in \cite{cluver2014}.

Of the 59 VGS galaxies, 36 have been detected in 12 $\rm{\mu}m$ and 18 have been detected in 22 $\rm{\mu}m$ 
with the WISE. We used the following zero magnitude flux densities which correspond to WISE bands (W1, W2, W3 and W4) with effective wavelengths 3.35 $\rm{\mu}m$, 4.60 $\rm{\mu}m$, 11.56 $\rm{\mu}m$ and 22.8 $\rm{\mu}m$ respectively: 309.68 Jy, 170.66 Jy, 29.05 Jy and 7.871 Jy   \citep{wright2010,jarrett2011,brown2014b}.

We use the calibrated 12 $\rm{\mu}m$ and 22 $\rm{\mu}m$ bands to derive the star formation 
rates \citep{jarrett2013} from warm-dust emission of the galaxies and compare them to $\rm{SFR\alpha}$ and 
$\rm{SFR_{NUV}}$ as described in section~\ref{sec:sfr}. We determine stellar mass from 3.4 $\rm{\mu}m$ 
and 4.6 $\rm{\mu}m$ bands according to
\cite{cluver2014} 
(section~\ref{sec:wise_sm}). The stellar mass data have been used to measure specific star formation 
rates (S\_SFRs), by normalising the $\rm{H\alpha}$ and UV SFRs by the 
stellar masses.

\section{Comparison samples}

For a proper interpretation of our results it is necessary to compare the star formation properties of the VGS galaxies with those of a comparison sample. It is difficult, however, to find 
such a sample because one needs galaxies which belong to environments that are only moderately denser than the voids and for which reliable data of similar quality is available. 
In addition, the environment for such a sample should be defined the same way as for the VGS sample. In very general terms 'the field' is everywhere but in void interiors, and
more than a several tidal radii away from dense clusters of galaxies. We call these environments 'the field' in this paper, they are the moderate density areas corresponding to the 
outer boundary regions of voids, the outskirts of filaments and the intermediate wall-like regions in the Cosmic Web \citep{aragon2010b,marius2014}. Finding a comparison sample that fulfils 
these requirements and 
for which data on the $\rm{H\textsc{i}}$, stellar mass and star formation properties, is available is not trivial. Our comparison samples consist of galaxies within the same stellar mass 
ranges as 
the VGS galaxies ($\rm{M_{*} <}$ $\rm{3 \times 10^{10}}$ $\rm{M_{\odot}}$). The stellar mass of the comparison samples is 
derived from infrared measurements of various bands, such as 2.2 $\rm{\mu}$m (K-band), WISE 3.4 $\rm{\mu}$m and Spitzer 3.6 $\rm{\mu}$m.  To be consistent with our data we made sure
that the $\rm{H\alpha}$ SFRs of the comparison sample are derived via $\rm{H\alpha}$ imaging, not from spectra.
They have been selected from three studies as described in the next subsections. One sample is 
complete and volume limited, the other samples are flux limited, but over a much larger volume.
These are subject to similar selection effects as the VGS sample.  

\subsection{A complete sample: Local volume field and isolated galaxies}

The advantage of a complete, volume limited sample is that it has a full inventory of all galaxies. 
If the volume is large enough that a range of environments is sampled, then a complete sample offers
the additional advantage that it is possible to compare the properties of galaxies in the different
environments. Such a complete sample has been put together for the Local Volume (LV) by
\cite{karac2013} (also see \cite{kara2013}). The LV galaxy catalogue includes 873
galaxies that are within 11 Mpc around the Milky Way or have corrected radial velocities
$\rm{v_{LG}}$ $<$ 600 $\rm{km}$ $s^-1$. The sample consist of many dwarf galaxies 
with $\rm{M_{B}}$ between -10 and -15 magnitudes. \cite{karac2013} define a parameter called 
'the tidal index $\rm{\Theta}$', 
that they use as a quantitative indicator for the density of a galaxy's environment. Using 
their classification, we have selected the galaxies which are either in the field of 
the local volume and/or isolated. Besides the $\rm{H\alpha}$ star formation information,
$\rm{H\textsc{i}}$ and stellar mass and $\rm{M_{B}}$ data are taken from their catalogue and atlas
of local volume galaxies. Stellar masses of these LV galaxies have been derived from their K-band 
luminosities. Within this LV galaxy catalogue, we have selected 115 local volume field and
 61 local volume isolated galaxies which are at the same stellar mass range as the VGS 
galaxies. 

\subsection{Flux limited samples: ALFALFA and JCMT NGLS}

The disadvantage of the volume limited LV sample is that is does not have many objects in the high 
stellar and $\rm{H\textsc{i}}$ mass range. To sample those one has to resort to flux limited samples
which cover a much larger volume, but are incomplete in the low mass range. Such samples do have
selection effects, but these are not dissimilar from those of the VGS sample, which is also biased
by the flux limit for spectroscopic data in SDSS. 

One flux limited comparison sample has been selected from the catalogue presented in 
\cite{gavazzi2012}. 
This catalogue results from an $\rm{H\alpha}$ imaging survey of galaxies drawn from the 
Arecibo Legacy Fast ALFA (ALFALFA) blind $\rm{H\textsc{i}}$ survey.
It consists of $\sim 400$ galaxies in the Local Supercluster and Virgo cluster. A subsequent
study \citep{gavazzi2013} presents the star formation properties of galaxies in the Virgo cluster 
and galaxies located in a large area around the Virgo cluster. 

Using the latter, we have 
created a comparison sample (ALFALFA low density) to compare, in particular, with the star formation properties 
of the VGS galaxies. Out of the 235 galaxies provided by \cite{gavazzi2012} we 
have selected galaxies which reside in underdense environments. For the selection we used the local 
environment measures based on volume-limited samples of ALFALFA galaxies selected 
from the SDSS spectroscopic survey \citep{jones2015}. The volume-limited sample covers the distance range $\rm{500 - 15,500}$ km $s^{-1}$/$H_{0}$ and 
includes all galaxies brighter than $\rm{M_{r} = - 18.9}$ mag. 
Two kinds of environment measures are calculated based on this volume-limited sample; fixed aperture 
and nearest neighbour distance.
In this study we adopt the environment measure based on the latter where the nearest neighbour density for each ALFALFA galaxy is calculated based on the projected distance to the third closest galaxy in the reference
SDSS catalogue. Using this
nearest neighbour statistics we selected the galaxies which have 3rd nearest neighbour density smaller than the mean 
density of -0.5 of the whole volume-limited sample (see Fig. 3 in \cite{jones2015}). 
\begin{figure}
 \centering\includegraphics[scale=0.3]{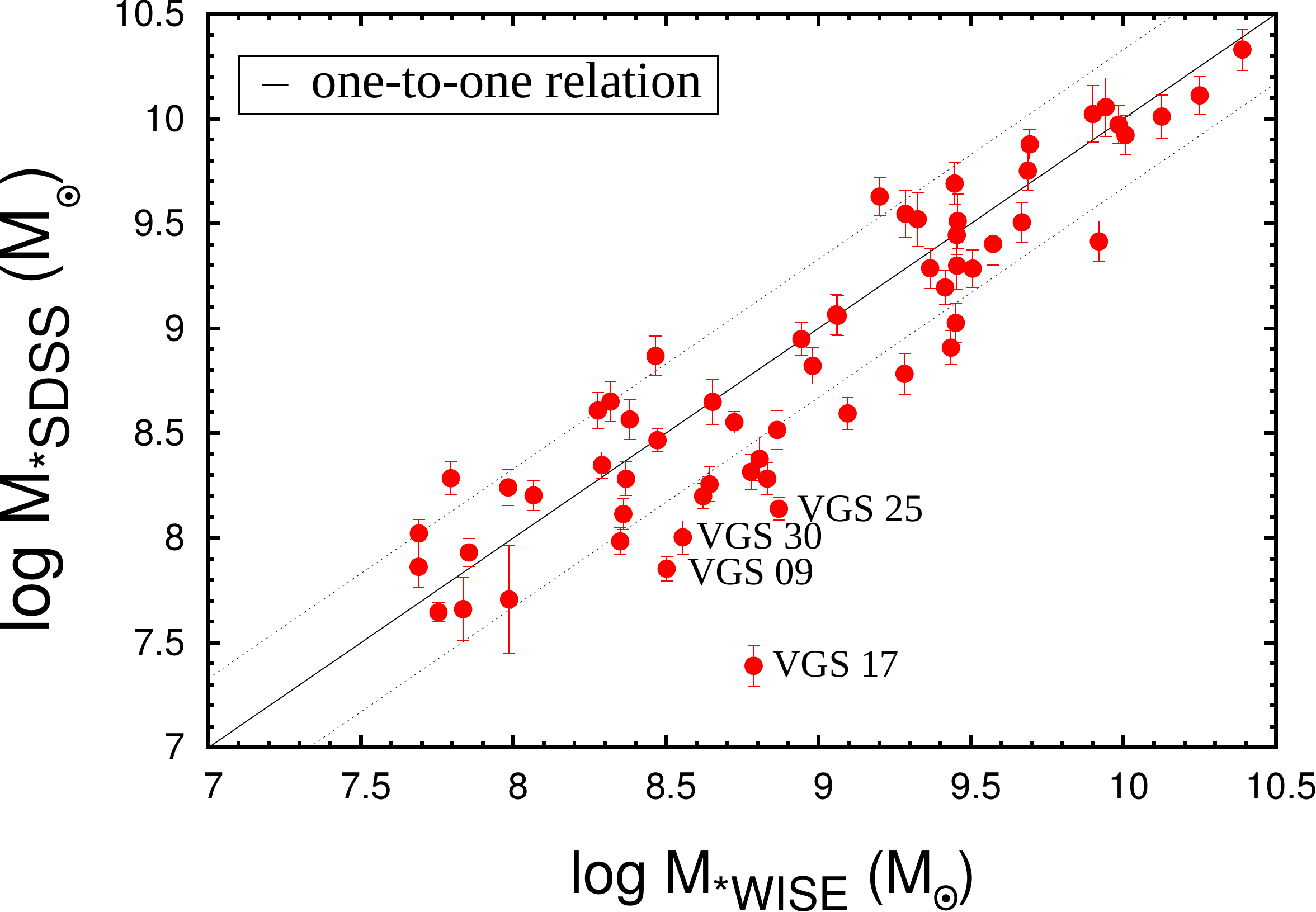}
 \caption{Comparison between the WISE-derived and SDSS MPA/JHU stellar mass estimates for the VGS galaxies. Solid black line indicates a one-to-one correspondence while 
 dotted lines are $\rm{1-\sigma}$ scatter. Outliers with low signal-to-noise in WISE are labelled.}
 \label{fig:mass_comp}
\end{figure}
After this we compare the entries 
with the Spitzer Survey of Stellar Structure in Galaxies ($\rm{S^{4}G}$) catalogue \citep{sheth2010,quer2015}. 
$\rm{S^{4}G}$ is a volume, magnitude and size limited 
($\rm{d < 40 Mpc}$, $\rm{|b| > 30^{\small o} }$, $\rm{m_{B} < 15.5}$, and $\rm{D_{25} > 1'}$) survey of 
2331 galaxies using the 3.6 and 4.5 $\rm{\mu m}$ fluxes from the 
Infrared Array Camera (IRAC) to derive stellar masses. The final ALFALFA comparison sample consists of 
33 galaxies. We use the SFRs and $\rm{H\textsc{i}}$ masses of these  galaxies as given in \cite{gavazzi2012}.

Another comparison sample was drawn from the
JCMT NGLS (Nearby Galaxies Legacy Survey), a project geared to imaging 156 nearby galaxies in CO and the 
NIR \citep{wilson2009}. \cite{sanchez2012} have defined 72 $\rm{H\textsc{i}}$ flux limited 
field galaxies as part of this survey. Among these, 43 have similar stellar mass as our VGS sample 
and these are added to our comparison sample. Their stellar masses have been taken from 
$\rm{S^{4}G}$ catalogue, the same way as done for the ALFALFA sample.

\section{Derived properties}
\subsection{Stellar masses}
\label{sec:wise_sm}

There are different methods to determine stellar masses. All are based on observations in different 
bands in the optical and the IR, and involve an estimate of the spectral energy distribution
coupled to stellar population models. For this study it is important that the stellar masses for 
both the VGS sample and the comparison samples have been determined with the same method to avoid
systematic effects. For this study we adopt to determine stellar masses using the
WISE [3.4]-[4.6] colour and the mass-to-light ratio relation of 3.4 $\rm{\mu}$m luminosity 
given in \citep{cluver2014}\footnote{WISE masses are calibrated using 
Galaxy and Mass Assembly (GAMA)-derived masses, which are
essentially SDSS colours.}. The assumption is that
the infrared 3.4 $\rm{\mu}$m and 3.6 $\rm{\mu}$m emission from galaxies mainly traces the old 
star population and has been shown to be an effective 
measure of galaxy stellar mass \citep{jarrett2013,meidt2014,cluver2014}. 
For low redshift sources this gives: 

\begin{equation}
 log_{10}M_{*}/L_{3.4} = -2.54 \times ([3.4]-[4.6]) -0.17
\end{equation}
where $L_{3.4}(L_{\odot}) = 10^{-0.4(M-M_{\odot})}$, $M$ is the absolute magnitude of the source in 3.4 $\rm{\mu}$m and $M_{\odot}=3.24$ \citep{jarrett2013}.

In Figure~\ref{fig:mass_comp} we compare our stellar mass estimates derived using WISE data to those of SDSS MPA/JHU catalogue for SDSS DR7 which we 
used in our earlier VGS studies (\cite{kreckel2012, kreckel2013}). 
The mass
estimates agree with a standard deviation of a factor of $\sim$ 2. Both SDSS and WISE stellar mass estimates agree on the upper limit of stellar mass of 
$\rm{M_{*} < 3 \times 10^{10}}$ $\rm{M_{\odot}}$ Outliers such as VGS\_09, VGS\_17, VGS\_25 and VGS\_30 have low signal-to-noise ($<$3), notably in [3.6]-[4.6] colour and are not
included into the specific star formation rate calculations.

\subsection{Star formation rates}
\label{sec:sfr}

Star formation rates can be estimated from $\rm{H\alpha}$ fluxes and from near UV fluxes. Both require a correction for extinction. An alternative way to
determine star formation rates is to use mid IR data, in particular emission at 
12 $\rm{\mu}m$ and 22 $\rm{\mu}m$ \citep{jarrett2013}. Below we describe the various methods 
and compare them.

\subsubsection{The $\rm{H\alpha}$ star formation rate}

Before calculating star formation rates one needs to correct the $\rm{H\alpha}$ fluxes for extinction. A well known method is to use the 
Balmer decrement from spectroscopic data.
We checked the Balmer decrements using the $\rm{H_{\alpha}}/\rm{ H_{\beta}}$ from the 
MPA-JHU DR7 catalogue for the SDSS 3$ ''$ spectra. We selected the good S/N spectra and we made sure that the SDSS fiber was at the correct place, by 
comparing the $\rm{H\alpha}$ flux with our own $\rm{H\alpha}$ images. Of the 75 VGS galaxies including companions, 46 of them there are 14 VGS galaxies whose optical and/or 
$\rm{H\alpha}$ disk sizes are comparable to the SDSS 3$ ''$ fiber aperture. 

For these galaxies SDSS spectra are representative for the whole galaxy disk and therefore the measured Balmer decrements can be safely used to correct the internal extinction.
If the Balmer decrement values are higher than the theoretical prediction, i.e. $\simeq$ 2.88, for T $\sim$ $10^{4}$K and an electron density of
 $\rm{n_{e} < 10^{6}}$ $\rm{cm^{-3}}$ \citep{calzetti2001}, we used its value to estimate the attenuation. Our sample has Balmer decrement values ranging from 2.9 to 6.1, for the objects 
 with $\rm{H_{\alpha}}/\rm{ H_{\beta}}$ > 2.88.
For the rest of the galaxies, we also looked at the 22 $\mu$m detections from the Wide-field Infrared Survey Explorer (WISE) and
the 4000-$\rm{\AA}$ break ($\rm{D_{n}(4000)}$) measurements from the SDSS spectra. If a galaxy has an $\rm{H\alpha}$ disk much larger than the 3$ ''$ aperture, a Balmer decrement value
 higher than 2.88 and is detected
 in 22 $\mu$, then we used the Balmer decrement value to estimate the extinction. This correction assumes that the 
$\rm{H_{\alpha}}/\rm{ H_{\beta}}$ is the same throughout the entire $\rm{H\alpha}$ emission region. If it is neither detected in 22 $\mu$ nor has a good quality SDSS spectra, 
then we used the near-UV extinction ($\rm{A_{NUV}}$), 
derived using the $\rm{D_{n}(4000)}$ parameter and k corrected $\rm{^{0.1}(NUV-r)}$ colour as described in the next subsection, to estimate the $\rm{H\alpha}$ extinction ($\rm{A_{\alpha}}$). 
For these cases we consider $\rm{D_{n}(4000)}$ measured in the 3$''$ fiber to be more representative for the entire galaxy than the Balmer decrement value.

In order to calculate the intrinsic $\rm{H\alpha}$ flux we followed the recipes in ~\cite{calzetti2000} and ~\cite{dominguez2012} which give:

\begin{equation}
I(H_{\lambda})= F(H_{\lambda}) \times 10^{(0.4 \times A_{\lambda})},
\end{equation}

where $I(H_{\lambda})$ and $F(H_{\lambda})$, are the intrinsic and the observed flux densities respectively and $\rm{A_{\lambda}}$ is the attenuation at wavelength
$\rm{\lambda}$ (in this case it is the attenuation in $\rm{H\alpha}$ ($\rm{A_{\alpha}}$). We have derived $\rm{A_{\alpha}}$ from the Balmer decrement ($\rm{f_{\alpha}/f_{\beta}}$) 
as stated in \cite{lee2009} or from $\rm{D_{n}(4000)}$ as outlined above.

\begin{equation}
A_{\alpha} = 5.91 \times log( \dfrac{f_{\alpha}}{f_{\beta}} ) - 2.7, 
\end{equation}

Star formation rates have been calculated from the $\rm{H\alpha}$ emission following the conversion of \cite{kennicutt2008}:

\begin{equation} 
SFR _{\alpha} [M_{\odot} yr^{-1}] = 5.4 \times 10^{-42} \times L(H_{\alpha}), 
\end{equation}

where $L (H_{\alpha}) $ is the luminosity, calculated as
  
\begin{equation}
L(H_{\alpha}) = 4 \; \pi \; d^{2} (3.086 \times 10^{24})^{2} \; I(H_{\alpha}), 
\end{equation}

where $d$ the distance to the galaxy in Mpc derived using spectral redshifts from SDSS and $I(H_{\alpha})$ is the extinction corrected $\rm{H\alpha}$ flux. Our results and overall trends do not depend
significantly on whether the SFRs are corrected for extinction or not.
\begin{figure*}
 \centering\includegraphics[scale=0.45]{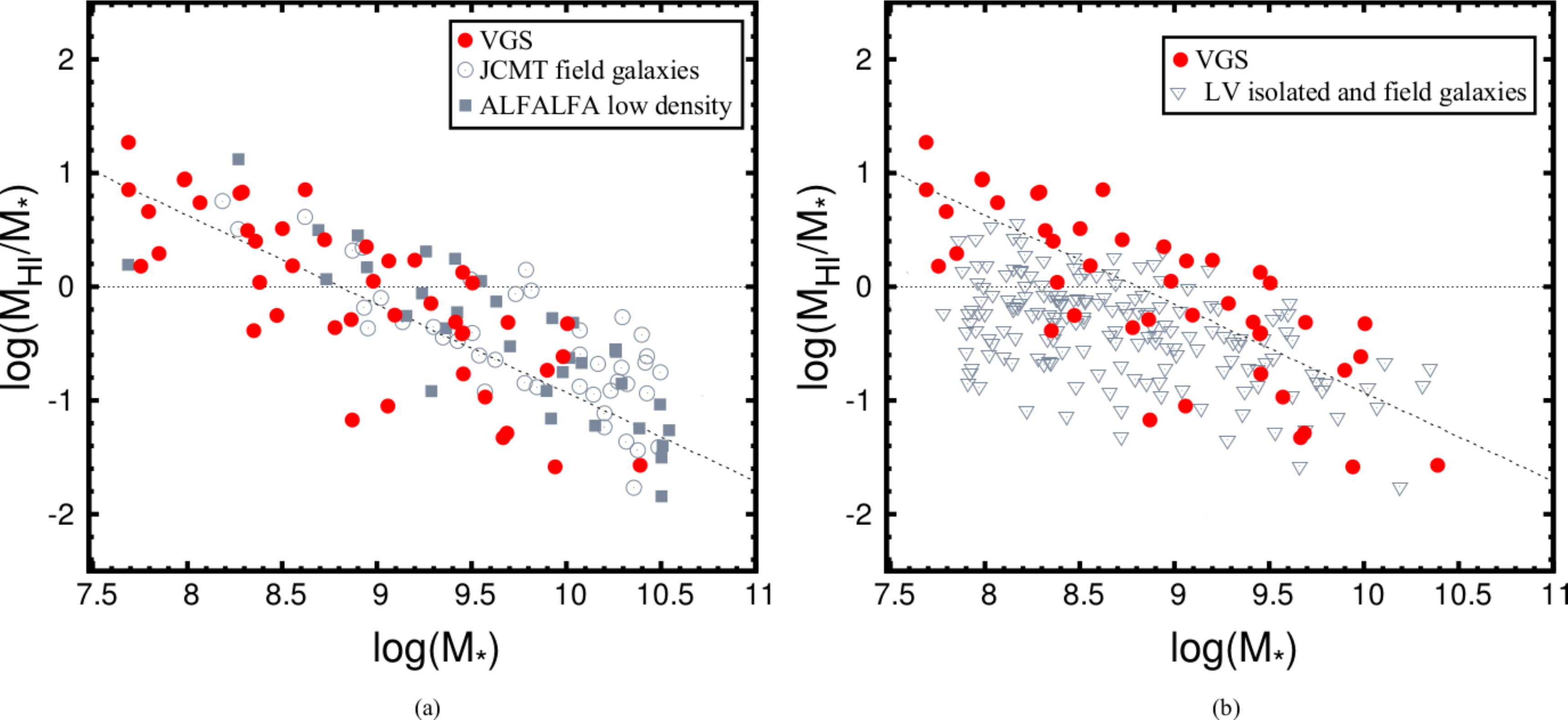}
 \caption{The $\rm{M_{H\textsc{i}}}$-to-$\rm{M_{*}}$
ratio for the VGS galaxies and galaxies from intermediate density environment as a function of stellar
mass. The comparison samples span the same magnitude range as the VGS galaxies. VGS galaxies are presented as filled red circles. 
The dashed line is the best fit ($\rm{log(M_{\textsc{Hi}}/M_{*})}$ = -0.78 $\times$ $\rm{log(M_{*})}$ + 6.87) to the VGS galaxies 
detected in $\rm{\textsc{Hi}}$. The dotted line indicates equal 
$\rm{{H\textsc{i}}}$ mass to stellar mass.
The comparison sample consist of ALFALFA galaxies of low density environments  
taken from \citet{gavazzi2012}, field galaxies 
(JCMT field galaxies) presented in \citet{sanchez2012} (a),  and local volume (LV) isolated and field galaxies defined in  
\citet{karachentsev2010,kara2013}, (b) LV galaxies are less $\rm{{H\textsc{i}}}$ rich for the given 
$\rm{M_{*}}$ range than
the VGS, ALFALFA and the JCMT field galaxies.}
\label{fig:3.4}
\end{figure*}
\subsubsection{The near UV star formation rate}

In order to estimate the star formation rate from near-UV, we first need to estimate the extinction $\rm{A_{NUV}}$ in the near-UV. This can be estimated using two methods. In the first method,  $\rm{A_{NUV}}$ is calculated using  $\rm{E(B-V)_{gas}}$. This has been derived in the previous subsection
for the objects where the Balmer decrement is representative for the whole $\rm{H\alpha}$ emitting disk. Following \cite{calzetti2001,kreckel2013}, this yields:

\begin{eqnarray}
A_{NUV} = 8.189 \times E(B-V)_{stars}, \\
E(B-V)_{stars} =  0.47 \times E(B-V)_{gas},
\end{eqnarray}
where $\rm{E(B-V)_{star}}$ is the colour excess of the stellar continuum.

The colour excess of the nebular emission lines, $\rm{E(B-V)_{gas}}$, has been calculated using $\rm{A_{\alpha}}$ according to \cite{cardelli1989}.

\begin{equation}
E(B-V)_{gas} =  \dfrac{A_{\alpha}}{2.532},
\end{equation}

We have used $\rm{E(B-V)_{gas}}$ to derive extinctions for the near-UV ($\rm{A_{NUV}}$), for the cases where we consider the Balmer decrement to be representative for the whole 
$\rm{H\alpha}$ emitting disk. 

In the second method, we use $\rm{D_{n}(4000)}$ and the (NUV-r) colour to calculate $\rm{A_{NUV}}$. SDSS \textit{r} magnitudes are corrected for galactic extinction, $\rm{ A_{\lambda}}$, 
using the catalogued extinctions \citep{schlegel1998}. For the near-UV magnitudes we calculated the galactic extinction following \cite{wyder2007}, who adopted $\rm{A_{r}/E(B-V) = 2.751}$ and 
$\rm{A_{NUV}/E(B-V) = 8.189}$, giving $\rm{A_{NUV} = 2.9807 A_{r}}$. 

The 4000-$\rm{\AA}$ break, $\rm{D_{n}(4000)}$ as defined in \cite{balogh1999}, is taken from the MPA/JHU catalogue.
If $\rm{D_{n}(4000) > 1.7}$, we adopt $\rm{A_{NUV}}$ = 0 (the only case is VGS\_05). If $\rm{D_{n}(4000) < 1.7 }$ then we follow 
\cite{calzetti2000} and use $\rm{A_{NUV}}$ = 0.81$\rm{A_{IRX}}$. $\rm{A_{IRX}}$ is defined in \cite{johnson2007} as:

\begin{equation}
A_{IRX} = 1.25 - 1.33x + 1.19y - 1.02xy,
\end{equation}
where $\rm{x = D_{n}}$(4000) - 1.25 and y = $\rm{^{0.1}(NUV-r)-2}$. We calculate the k-corrected NUV and r band
magnitudes band shifted to z=0.1, $\rm{^{0.1}(NUV-r)}$, using the kcorrect (v4.1.4) package using the method
of \cite{blanton2007}.

The $\rm{SFR_{NUV}}$ has been calculated from the GALEX near-UV luminosities, corrected for internal dust attenuation
following the method outlined in \cite{schiminovich2010}.

\begin{equation}
 SFR_{UV} [M_{\odot} yr^{-1}] =\frac{L_{UV}f_{UV}(young)10^{0.4A_{UV}}}{\eta_{UV}},
\end{equation}

where $\rm{L_{UV}}$ is the luminosity in $\rm{erg\; s^{-1} Hz^{-1}}$, $\rm{f_{UV} (young)}$ is the fraction of light that originates
in young stellar populations, $\rm{\eta_{UV}}$ is the conversion factor between UV luminosity and recent-past-
averaged star formation rate and $\rm{A_{UV}}$ is the dust attenuation. Following \cite{schiminovich2010}, we assumed $\rm{f_{UV} (young)}$ = 1 and $\rm{\eta_{UV}}$ = $\rm{10^{28.165}}$. 
\begin{figure}
\centering\includegraphics[scale=0.33]{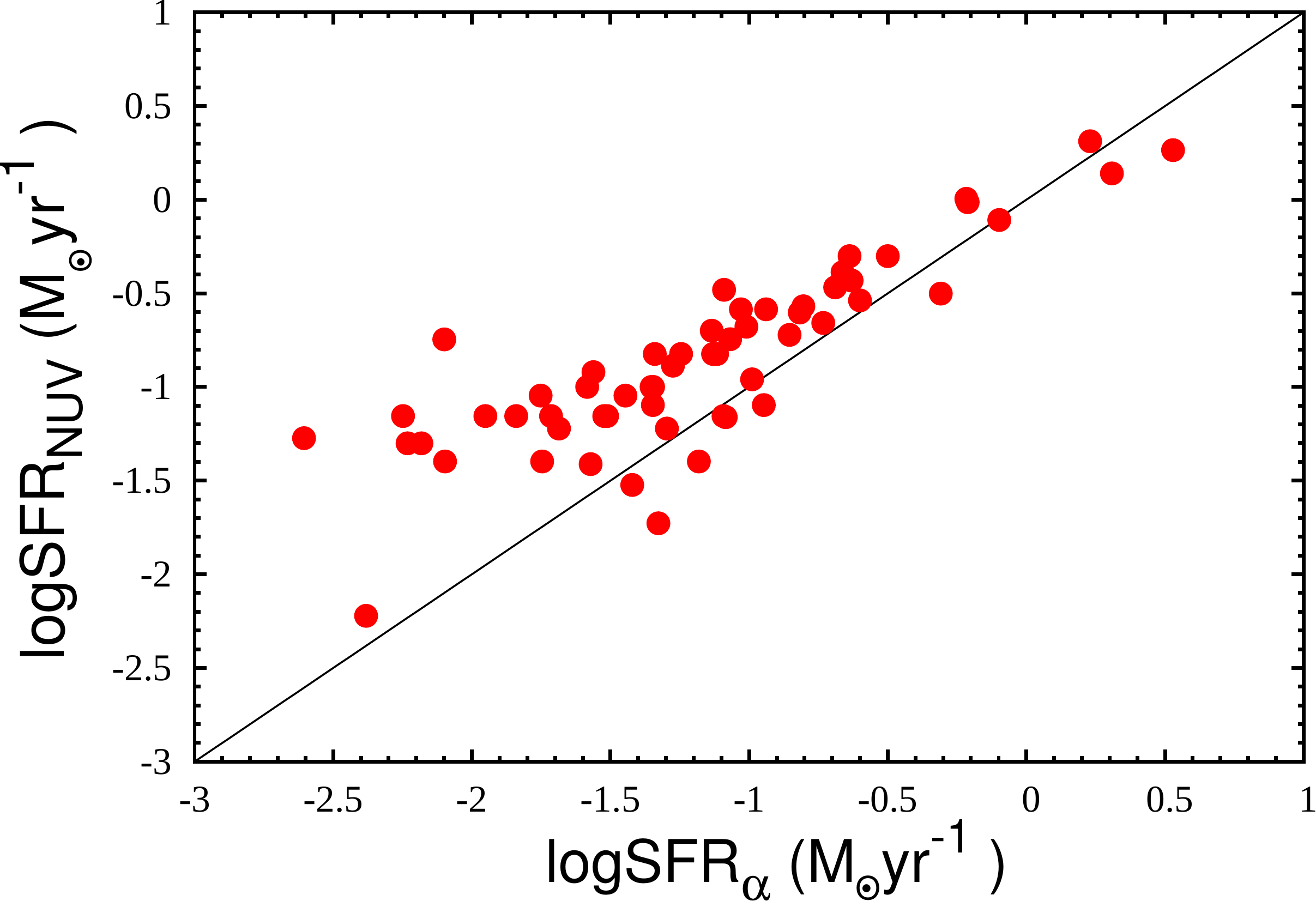} 
 \caption{$\rm{SFR\alpha}$ against $\rm{SFR_{NUV}}$ of the VGS galaxies.}
  \label{fig:alpha_nuv}
\end{figure}
Most of the companion galaxies do not have spectra in SDSS and thus do not have a measured
value for $\rm{D_{n}(4000)}$. This is problematic, as the dust attenuation in the near-UV is significant and
can decrease the observed SFR by up to an order of magnitude. For galaxies without a measured $\rm{D_{n}(4000)}$ we assume a fixed value of 1.25, which is
the median value from \cite{johnson2007} and appears appropriate for our stellar mass range (see
Figure 1, \cite{kauffmann2003}). A choice of $\rm{D_{n}(4000)}$=1.0 would increase the SFR by $\rm{10-30 \%} $.
\subsubsection{The mid IR star formation rate}
Star formation rates from the mid IR fluxes at 12 $\rm{\mu}m$ and 22 $\rm{\mu}m$
($\rm{SFR_{12}}$ and $\rm{SFR_{22}}$ respectively) have been calculated following the 
conversion of \cite{jarrett2013},
\begin{align}
 SFR_{12} (\pm 0.28)(M_{\odot} yr^{-1}) = 4.91(\pm0.39) \times 10^{-10} \nu L_{12}(L_{\odot}),\\
SFR_{22} (\pm 0.04)(M_{\odot} yr^{-1}) = 7.50(\pm0.07) \times 10^{-10} \nu L_{22}(L_{\odot}),
\end{align}
where $\rm{\nu L_{12}}$ and $\rm{\nu L_{22}}$ are the luminosity densities for the 12 $\mu$m and 22$\mu$m and $\rm{L_{\odot}}$ is the total solar 
luminosity equal to $\rm{3.839 \times 10^{33}}$ erg $\rm{s^{-1}}$.
\begin{figure*}
\includegraphics[scale=0.44]{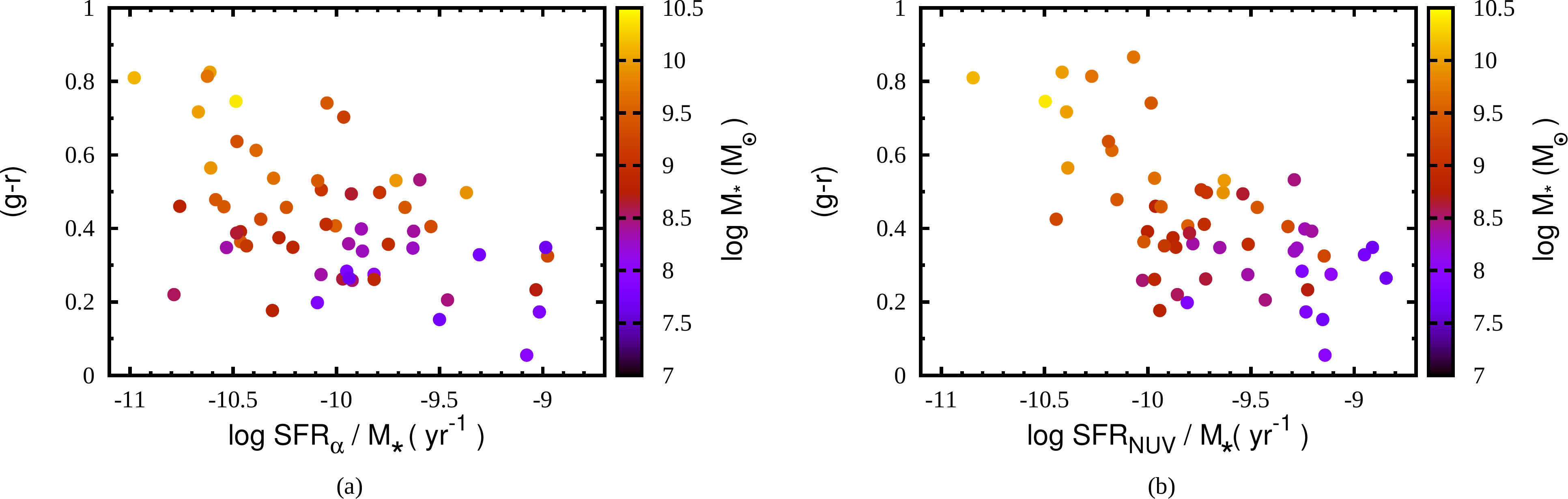} 
 \caption{Left: ($g-r$) colour versus $\rm{H\alpha}$ specific star formation rates ($\rm{SFR\alpha/M_{*}}$) of the VGS galaxies, colour coded as a function 
 of their stellar mass. 
Right: Similar to the previous plot except that ($g-r$) is plotted against the near UV specific star formation rates ($\rm{SFR_{UV}/M_{*}}$). There is a tighter correlation compared to the
$\rm{H\alpha}$ case. }
  \label{fig:3.7}
\end{figure*}
\section{Results} 
By combining the star formation rates with 
$\rm{H\textsc{i}}$ and stellar masses, we can examine the general star formation properties, specific star formation rates and
star formation efficiencies of our void galaxy sample and compare these with those of field galaxies. Our results, showing the basic scaling relations, are presented in 
the following subsections. We examine the star formation, stellar and $\rm{H\textsc{i}}$ 
properties of the VGS galaxies and compare these with the same properties
of objects in the comparison samples.  
\subsection{$\rm{H\small\textsc{I}}$ and stellar masses}
In Figure~\ref{fig:3.4} we show the $\rm{M_{H\textsc{i}}}$-to-$\rm{M_{*}}$ 
ratios (gas fractions) as a function of $\rm{M_{*}}$ of the VGS galaxies 
and the galaxies in the comparison samples.  
For a clear presentation, we separate the ALFALFA and JCMT samples (Figure~\ref{fig:3.4}a) from the LV galaxy sample 
(Figure~\ref{fig:3.4}b). In both plots, 
we show the best fit, $\rm{log(M_{HI}/M_{*})}$ = -0.78 $\times$ $\rm{log(M_{*})}$ + 6.87, to the VGS galaxies to highlight the distribution of the 
comparison sample with respect to the VGS galaxies. 
 \begin{figure}
  \centering\includegraphics[scale = 0.32]{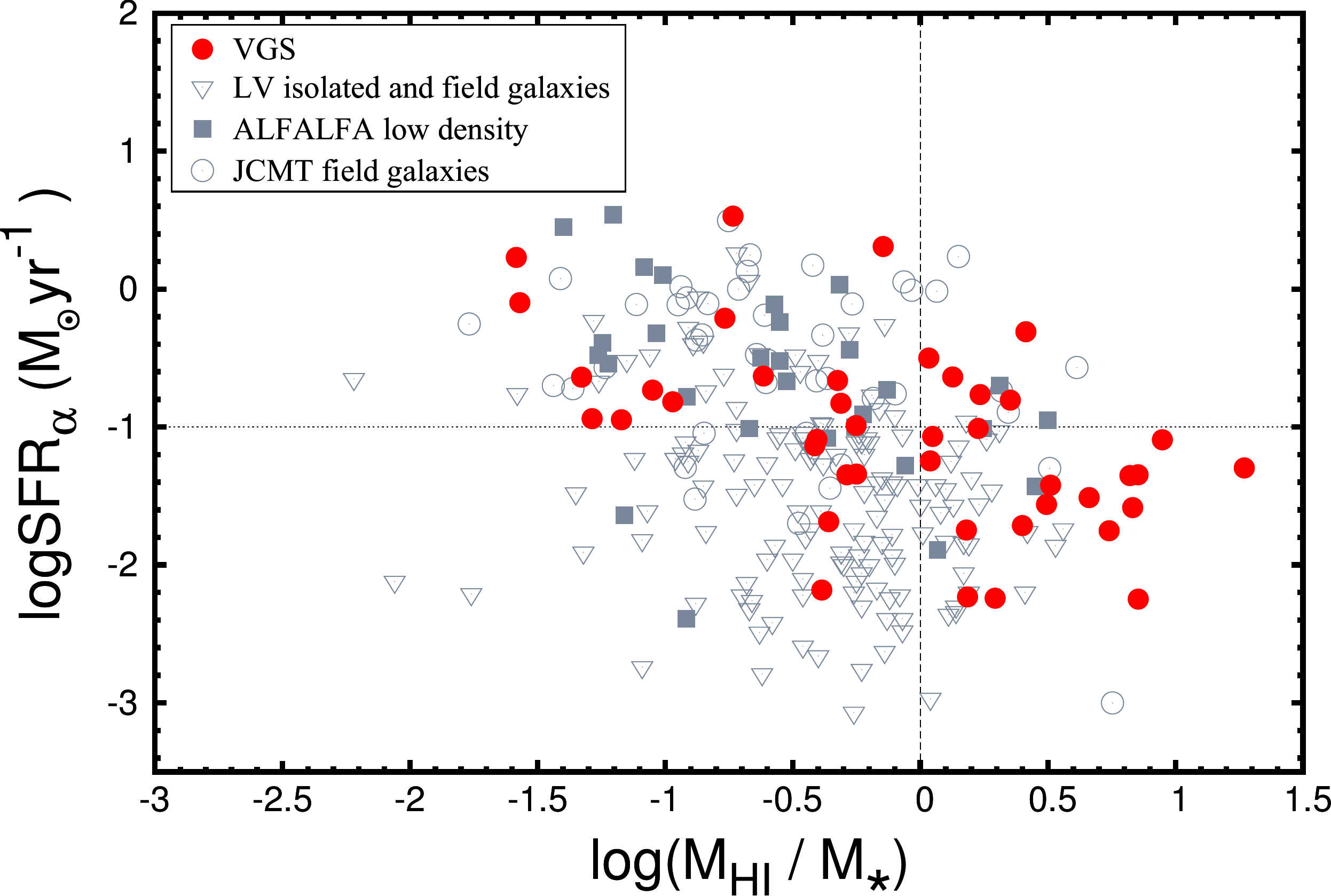}
 \caption{The extinction corrected $\rm{SFR\alpha}$ plotted against the $\rm{M_{HI}}$-to-$\rm{M_{*}}$ ratios for the VGS galaxies and the comparison sample.}
\label{figure:13}
\end{figure}
The VGS galaxies are very similar in gas content to the ALFALFA/JCMT galaxies.
In the lower stellar mass range ($M_{*} < 10^{9}$) the comparison with the
ALFALFA/JCMT sample is difficult because of a lack of objects. There a 
comparison with the LV sample is more appropriate. In this mass range the
VGS galaxies tend to be more gas rich then their LV counterparts, giving the
general impression that in particular toward lower stellar masses the VGS 
galaxies are gas rich systems. 
\subsection{Current and recent star formation properties}
There are many studies in the literature comparing the two star formation tracers $\rm{H\alpha}$ and near-UV.
 Such a study, however, has never been specifically carried out for void galaxies. Here,
we interpret our results  using previous studies which have investigated both these two regimes. First we compare the two star formation tracers in 
Figure~\ref{fig:alpha_nuv}, then we examine 
 the $\rm{H\alpha}$ and near-UV 
star formation rates of the VGS galaxies as a function of $g-r$ colour and stellar mass. For each relation, we present these scaling relations 
derived from the $\rm{H\alpha}$ and the near-UV fluxes after a correction for extinction\footnote{As it will be discussed in the following sections, our sample consists of an AGN.However, it is not plotted,
since we can not distinguish the AGN  from the $\rm{H\alpha}$ emitting disk.} In general, we see similar trends to those 
observed in previous studies. 
\begin{figure*}
\centering
\subfigure[]
{
 \includegraphics[scale=0.32]{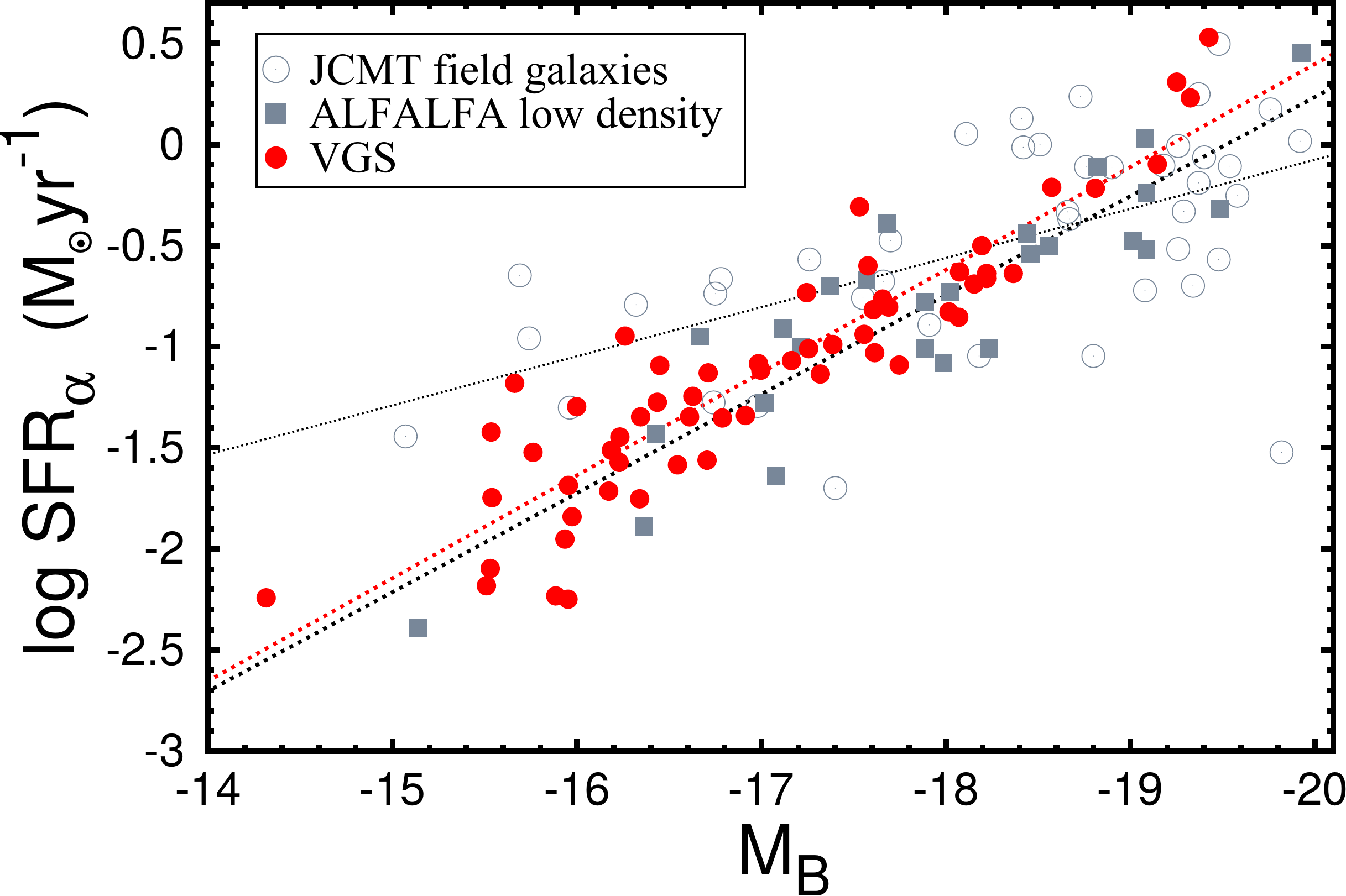} 
\label{}
}
\subfigure[]
{
 \includegraphics[scale=0.32]{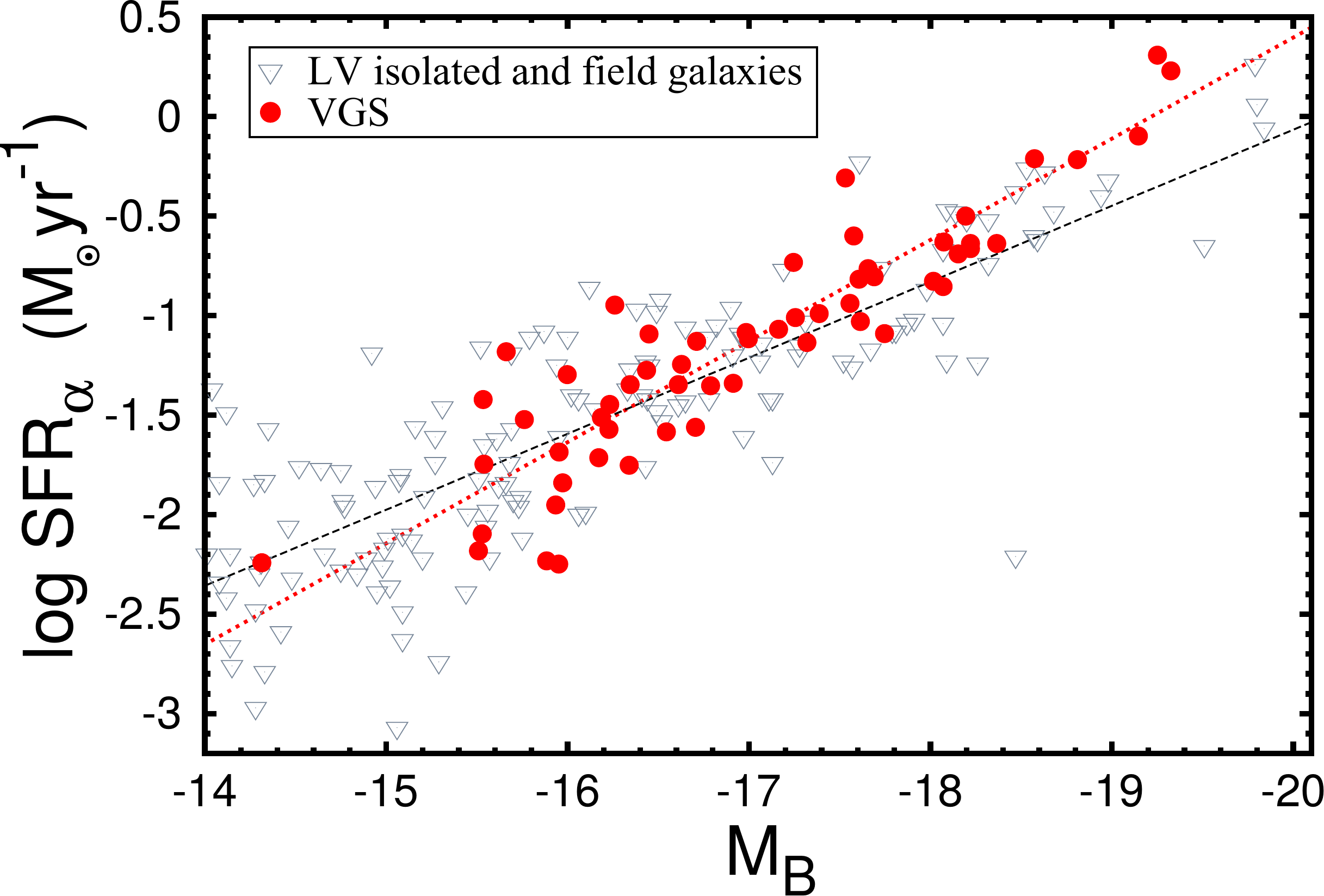} 
\label{}
}
\caption{ (a) and (b): The extinction corrected $\rm{SFR\alpha}$ against $\rm{M_{B}}$ for the VGS galaxies and the comparison sample. Best fits 
for the 
VGS galaxies ($\rm{logSFR\alpha}$ = -0.5 $\times$ $\rm{M_{B}}$ - 9.77), JCMT field galaxies 
($\rm{logSFR\alpha}$ = -0.25 $\times$ $\rm{M_{B}}$ - 5), 
ALFALFA Virgo surrounding 
($\rm{logSFR\alpha}$ = -0.49 $\times$ $\rm{M_{B}}$ - 9.56) and LV samples 
($\rm{logSFR\alpha}$ = -0.4 $\times$ $\rm{M_{B}}$ - 7.95) are indicated with the dashed lines. }
\label{fig:3.5}
\end{figure*}
According to Figure~\ref{fig:alpha_nuv}, $\rm{SFR\alpha}$ is systematically lower than $\rm{SFR_{NUV}}$ at star formation rates below 
$\rm{SFR\alpha}$ $\sim$ 0.03 $\rm{M_{\odot}}$ $\rm{yr^{-1}}$. This discrepancy between $\rm{H\alpha}$ and UV star formation rates
has been investigated in several studies 
\citep{sullivan2000,bell2001,buat2002,lee2009,lee2011,kara2013}. In most cases,
they compared $\rm{H\alpha}$ to far UV, instead of to near-UV. The main result is that, after applying proper corrections for dust 
attenuation, 
$\rm{H\alpha}$ tends to increasingly under-predict the total SFR relative to the far UV at the faint end of luminosity function 
($\rm{L_{\alpha} \leq 10^{37}}$ erg $\rm{s^{-1}}$). \cite{lee2009} showed that the average $\rm{H\alpha}$ to far UV 
flux ratio is lower than expected by a factor of two, at $\rm{SFR\alpha}$ $\sim$ 0.003 $\rm{M_{\odot}}$ $\rm{yr^{-1}}$. In our sample 
this decrease is much more prominent and steep and 
starts at around
$\rm{SFR\alpha}$ = 0.03 $\rm{M_{\odot}}$ $\rm{yr^{-1}}$. Several suggestions have been made to explain the observed trends in $\rm{SFR\alpha/SFR_{FUV}}$. These 
include: the effects of 
uncertainties in the stellar evolution tracks and model atmospheres, non-solar metallicities, non-constant SFHs, leakage of ionising
 photons, departures from Case B recombination, dust 
attenuation, stochasticity in the formation of high-mass stars and variations in the IMF \citep[see][for a review]{meurer2009,lee2009}. 
The VGS galaxy sample contains many low mass galaxies. In
particular the VGS galaxies with very low star formation rates have stellar masses around $\rm{10^{8}}$ $\rm{M_{\odot}}$ and lower, and as we will discuss in 
$\S$~\ref{sec:infrared}, the VGS galaxy sample consists of galaxies with quite a range in metallicity. 
So it is important to bear this difference between UV and $\rm{H\alpha}$ SFRs
in mind.
\begin{figure}
\centering
\subfigure[]
{
 \includegraphics[scale=0.12]{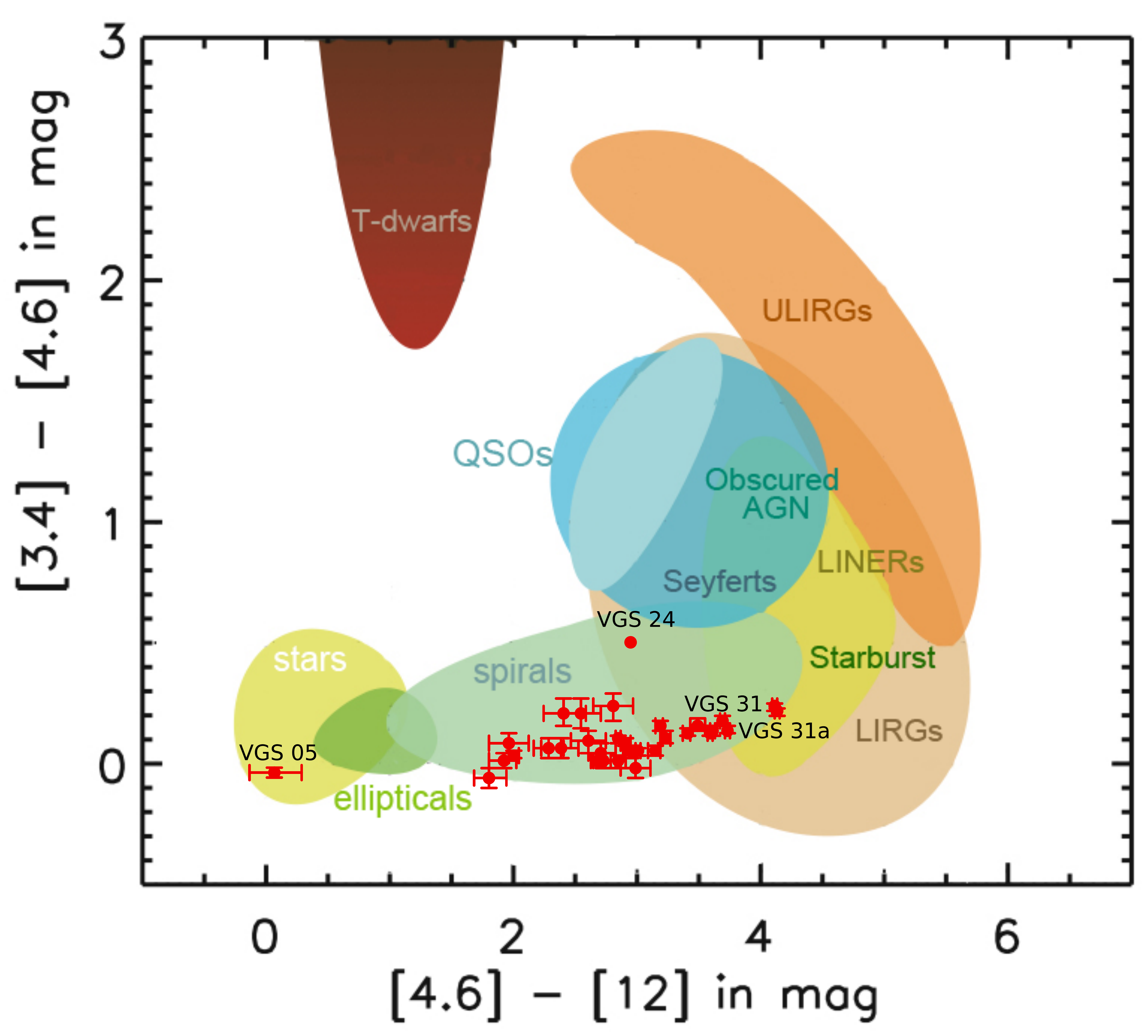} 
 \label{}
 }
 \subfigure[]
{
 \includegraphics[scale=0.33]{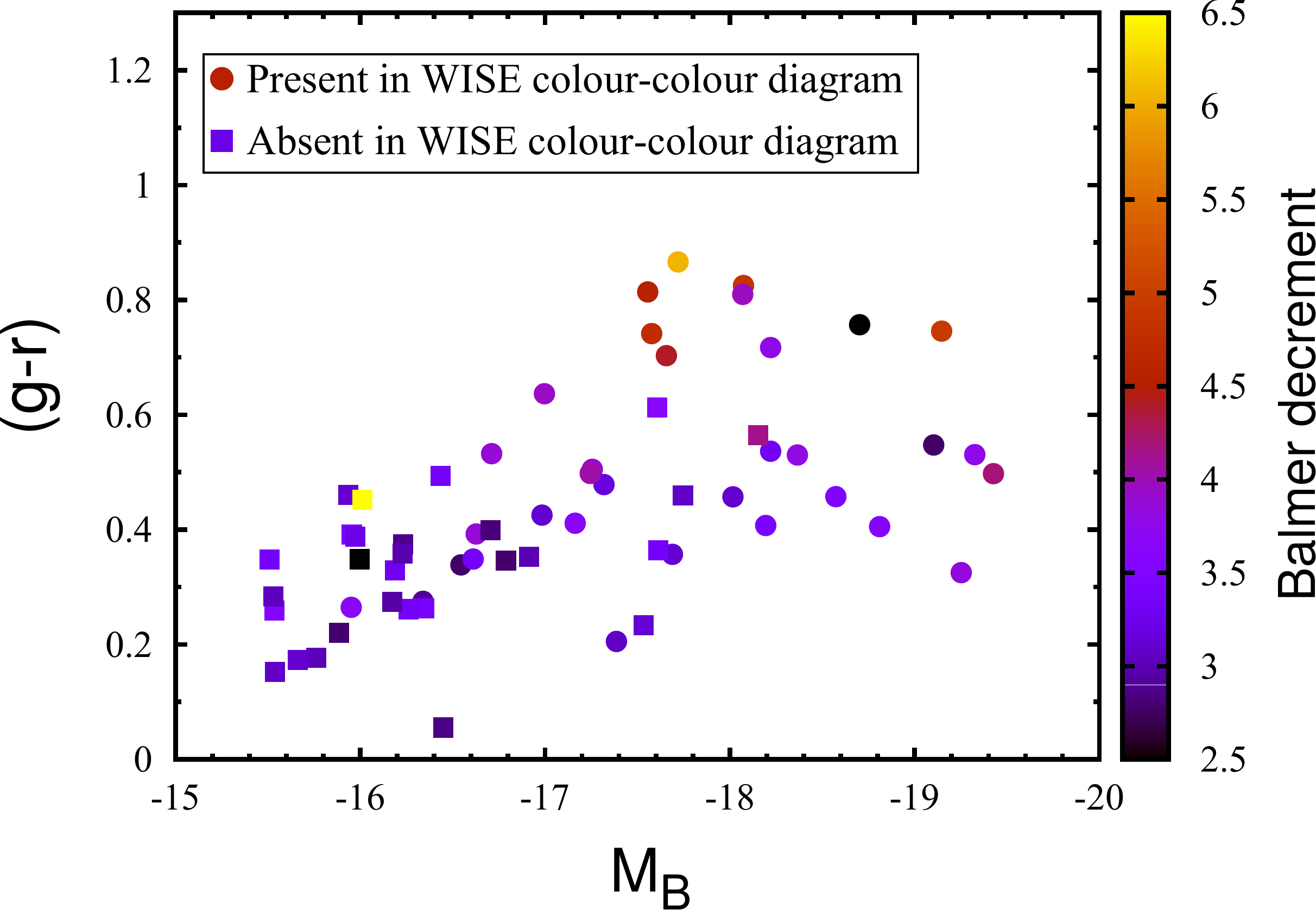}
 \label{}
}
  \caption{(a) VGS galaxies (red dots) overlaid on the WISE  [3.4]- [4.6] against [4.6]-[12] colour-colour diagram (without k-corrections) adapted from \citet{jarrett2011}. 
  Only the objects with magnitude error $<$ 0.2 mag in both colours are shown. Some of the particular examples in the VGS sample are indicated. VGS\_05;
   an early type galaxy, 
VGS\_24; which is the only AGN and Markarian galaxies VGS\_31 and 31a are labelled in the diagram as well. (b) Colour-magnitude diagram based on SDSS $g-r$ colour. 
Galaxies that are plotted in (a) shown as dots and galaxies that are absent in the colour-colour diagram are shown as squares. } 
  \label{figure:wise_bubble}
\end{figure}
In Figure~\ref{fig:3.7}, we compare ($g-r$) colour of the VGS galaxies with
 their specific star formation rates (SFR/$\rm{M_{*}}$), computed for both $\rm{H\alpha}$ (Figure~\ref{fig:3.7}a) and 
near-UV (Figure~\ref{fig:3.7}b). We add the stellar masses as a third parameter in this comparison and plot the VGS galaxies, colour coded as 
a function of their stellar mass. According to the colour-star formation-stellar mass relation, smaller galaxies are in general bluer and have higher specific star 
formation rates for a given mass range \citep{brinchmann2004,kauffmann2004}. VGS galaxies display the expected trend although there is a difference between the $\rm{H\alpha}$ and near-UV regimes. 
It is clear that this 
difference depends on stellar mass. For the $\rm{H\alpha}$, there is no single 
linear correlation for the whole stellar mass range (Figure~\ref{fig:3.7}a). Both low and high mass galaxies are present for a given $\rm{SFR\alpha/M_{*}}$ range.  
In the near-UV domain,
however, there is a tighter linear relation between colour and near-UV specific star formation rates (Figure~\ref{fig:3.7}b). Besides, 
galaxies are uniformly distributed in this linear relation with respect to their stellar mass; low stellar mass galaxies have higher specific 
star formation rates and are blue, higher stellar mass ones have lower specific star formation rates and are red. This supports the 
hypothesis that small galaxies suffer much more from stochasticity, affecting $\rm{H\alpha}$ samples more as $\rm{H\alpha}$ is more sensitive to the very  
recent star formation (time scale is $~5\times10^{6}$year for $\rm{H\alpha}$ and $~10^{7}$year for UV), as noted by \cite{lee2009}. 

The majority of the VGS galaxies have $\rm{SFR\alpha}$ less than
 1 $\rm{ M_{\odot}} yr^{-1}$. The mean $\rm{SFR\alpha}$ is $\sim$ 0.1$\pm$0.03 $\rm{M_{\odot}}$ $\rm{yr^{-1}}$ and  the luminosities range between 
$\rm{\sim 3 \times 10^{38}}$ $\rm{erg}$ $\rm{s^{-1} < \rm{L_{\alpha}} \lesssim 2.4 \times 10^{41}}$ $\rm{erg}$ $\rm{s^{-1}}$.  When corrected for the extinction, the mean 
$\rm{SFR\alpha}$ becomes $\sim$ 0.2 $\pm$ 0.06 $\rm{M_{\odot}}$ $\rm{yr^{-1}}$ and the luminosity increases to 6 $\times$ $10^{41}$ $\rm{erg}$ $\rm{s^{-1}}$. 

Similar to the $\rm{H\alpha}$ star formation rates, most of the galaxies have $\rm{SFR_{NUV}}$ below 1 $\rm{ M_{\odot}} yr^{-1}$. 
The mean  $\rm{SFR_{NUV}}$ of the VGS galaxies is  $\sim$ 0.08 $\rm{M_{\odot}}$ $\rm{yr^{-1}}$, and $\sim$ 0.35 $\rm{M_{\odot}}$ $\rm{yr^{-1}}$ when corrected for internal extinction.

In Figure~\ref{figure:13} we compare the $\rm{SFR\alpha}$ of the VGS and the comparison sample galaxies to their gas fractions $\rm{M_{HI}}$/$\rm{M_{*}}$.
The $\rm{SFR\alpha}$ of the VGS galaxies increases as their gas fractions decrease. This is true for the ALFALFA and the 
JCMT galaxies as well. LV galaxies, on the other hand, populate the area
with gas fractions below 1 and $\rm{SFR\alpha}$ $< 0.1\  \rm{M_{\odot}}$ $\rm{yr^{-1}}$. and do not show a significant trend in $\rm{SFR\alpha}$
with decreasing gas fraction.

When comparing the $\rm{SFR\alpha}$ to $\rm{M_{B}}$ (Figure~\ref{fig:3.5}a and b), both the VGS galaxies and the comparison sample display the 
well known 
star formation - luminosity correlation, with VGS galaxies having a slope of -0.5. This is steeper than the slope of -0.29 and -0.41 
found by \cite{sanchez2012} 
and \cite{lee2009b}, respectively. The ALFALFA sample has a very similar slope as the VGS galaxies, while the JCMT sample shows large scatter and a shallower slope.
\begin{figure*}
  \centering\includegraphics[scale = 0.45]{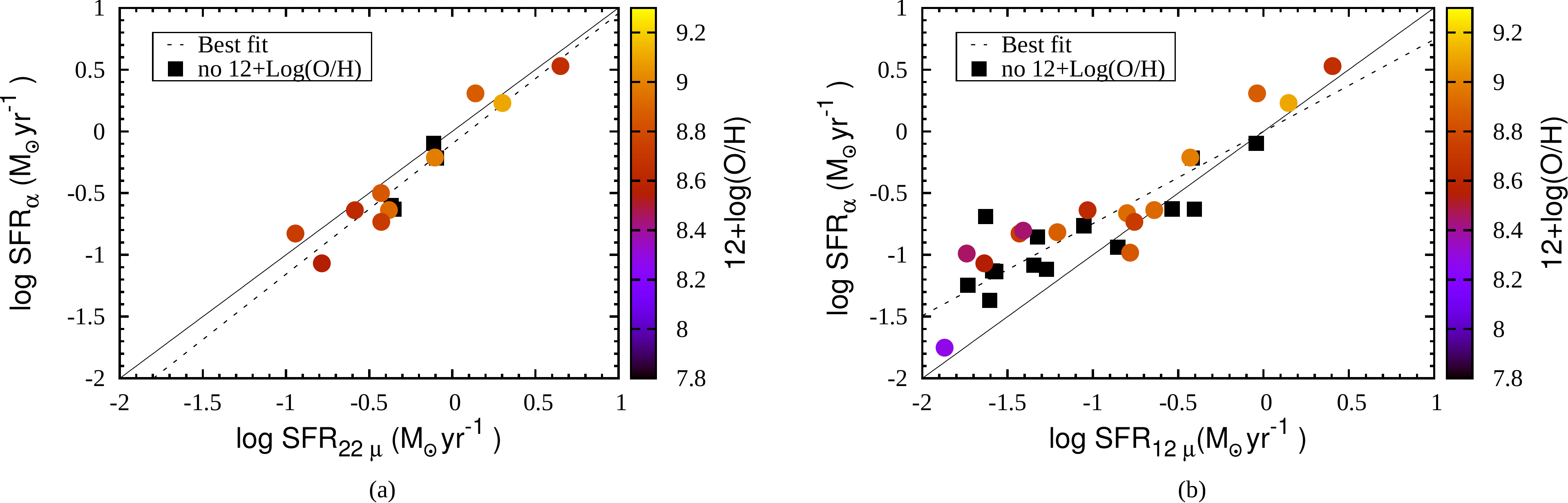}  
  \caption{(a): $\rm{SFR_{22}}$ against $\rm{SFR\alpha}$ as a function of gas-phase metallicity.. 
  (b): Same as the previous plot, except that $\rm{SFR_{12}}$ is plotted instead of $\rm{SFR_{22}}$. In both diagrams black lines correspond to a one-to-one correlation while 
  the dashed lines are the best fit. Black squares are the galaxies that are not present in the gas-phase metallicity catalogue of the MPA/JHU catalogue for
SDSS DR7.  }
\label{figure:sf_w4}
\end{figure*}

Many LV galaxies have much lower
luminosities than the VGS galaxies and they have a large spread in their SFRs at the low luminosity end (Figure~\ref{fig:3.5}b).
The same is true for the VGS galaxies although they do not probe as low luminosities as the LV galaxies do. 
There is only one VGS galaxy with $\rm{M_{B}}$ fainter than -15.

Despite the large range in luminosity covered by the VGS galaxies and the 
comparison samples and the scatter in the data, it is quite remarkable that 
all obey the star formation - luminosity correlation so well.

\subsection{Colour and star formation properties}
The IR data can be used to probe the old stellar population
as well as the more recent star formation. 
The WISE 3.4 $\rm{\mu}$m and 4.5 $\rm{\mu}$m bands are dominated by old stellar
populations and sensitive to hot dust whereas the 12 $\rm{\mu}$m band is sensitive to star formation  and dust continuum, and dominated by 11.3 $\rm{\mu}$m PAH emission. The 
colour-colour diagram of these bands has been shown to be an efficient way of separating old stellar population 
dominated systems from star-forming systems and AGNs \citep{jarrett2011,stern2012,cluver2014}. In Figure 
\ref{figure:wise_bubble}a we overplot VGS galaxies that have high S/N detections at 3.4 $\rm{\mu}$m and 4.5 $\rm{\mu}$m on the 
WISE colour-colour diagram of \cite{jarrett2011}. VGS galaxies lie on the green zone of 
spiral galaxies. Their colour $[4.6]-[12] > 1.5$ is consistent with actively star forming systems. Some of the peculiar examples of our sample are also
indicated on the plot. VGS\_31 and 31a are two Markarian galaxies and have the highest 
SFRs of the whole sample. They occupy a region overlapping with starbursts. The AGN VGS\_24 is clearly separate from the rest of the VGS galaxies and closer to the 
region of QSOs, Seyferts and AGNs. VGS\_05 is very blue (in [3.4] - [4.6] colour) in the infrared as most early type galaxies. 
Galaxies in the WISE colour-colour diagram are 
plotted together with the rest of the VGS galaxies in an optical colour-magnitude diagram in Figure \ref{figure:wise_bubble}b. In this diagram we show the relation 
between the SDSS $g-r$ colour, the absolute
 B-band magnitude and the Balmer decrement of the VGS galaxies. Galaxies plotted in Figure \ref{figure:wise_bubble}a are indicated as dots and the 
 remaining galaxies as squares. Most of the red galaxies in Figure \ref{figure:wise_bubble}b have higher Balmer decrement values 
 ($4 \leq H\alpha/H\beta \leq6$) and are also 
 detected at 3.4 $\rm{\mu}$m, 4.6 $\rm{\mu}$m and 12 $\rm{\mu}$m. This shows that these objects are red (in $g-r$ colour) due to extinction and they are not early types 
 galaxies that still form stars. 
 
The distribution of the VGS galaxies in the WISE colour-colour diagram and in the optical colour-magnitude diagram suggest that these objects 
 have a significant old stellar population as well as young population which is also found by \cite{penny2015}. On the other hand, blue and small galaxies in 
 Figure \ref{figure:wise_bubble}b (except a couple of objects),
 have smaller Balmer decrement values so they are not significantly affected by extinction. The fact that they are blue and not bright in WISE bands 
 show that these galaxies are dominated mostly by young stars. For fainter galaxies the metallicity is lower, so the extinction as well, therefore it is difficult to say
anything about the amount of old stars.
\begin{figure*}

\subfigure[]
{
 \includegraphics[scale=0.33]{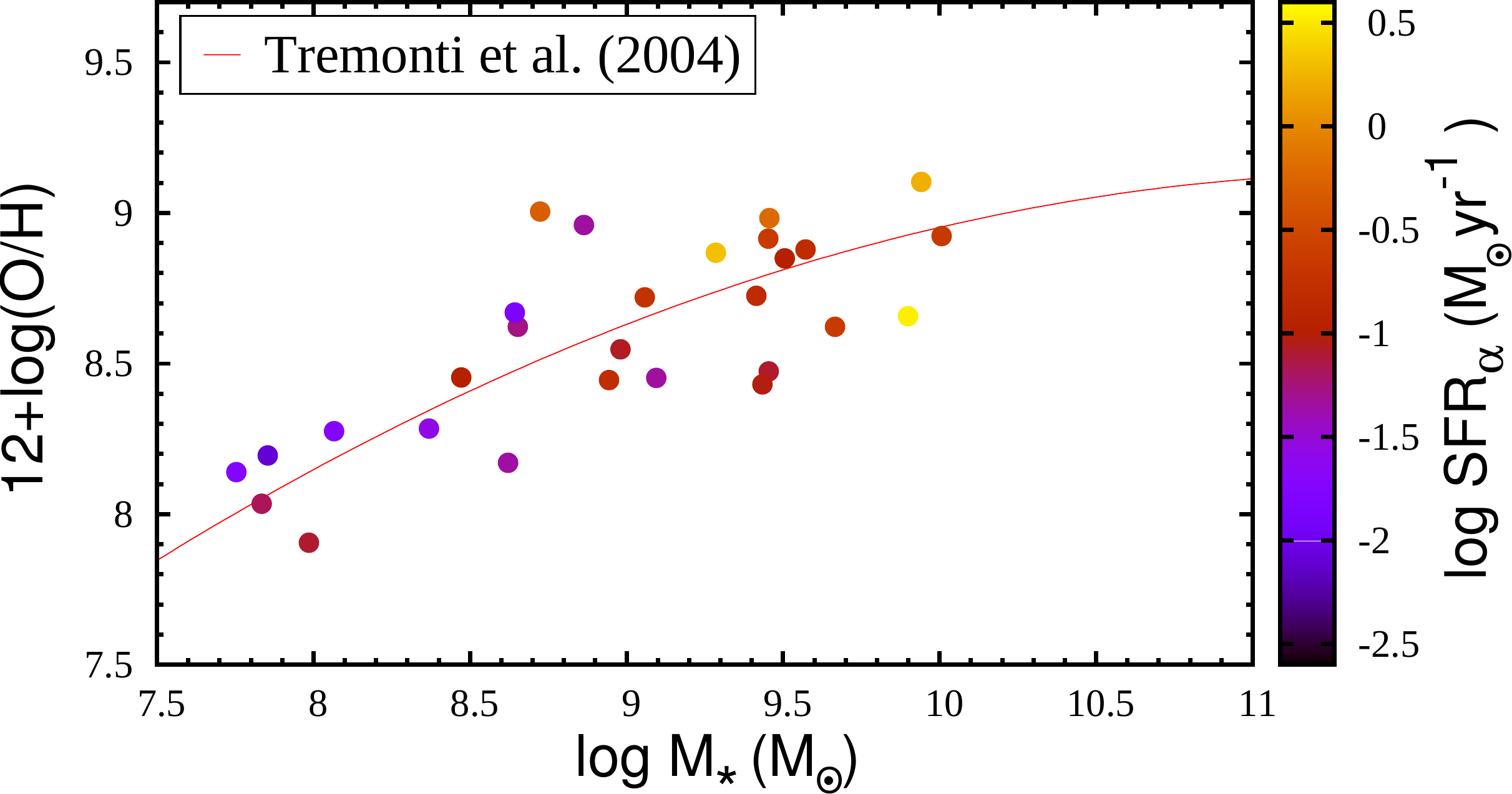} 
\label{}
}
\subfigure[]
{
 \includegraphics[scale=0.33]{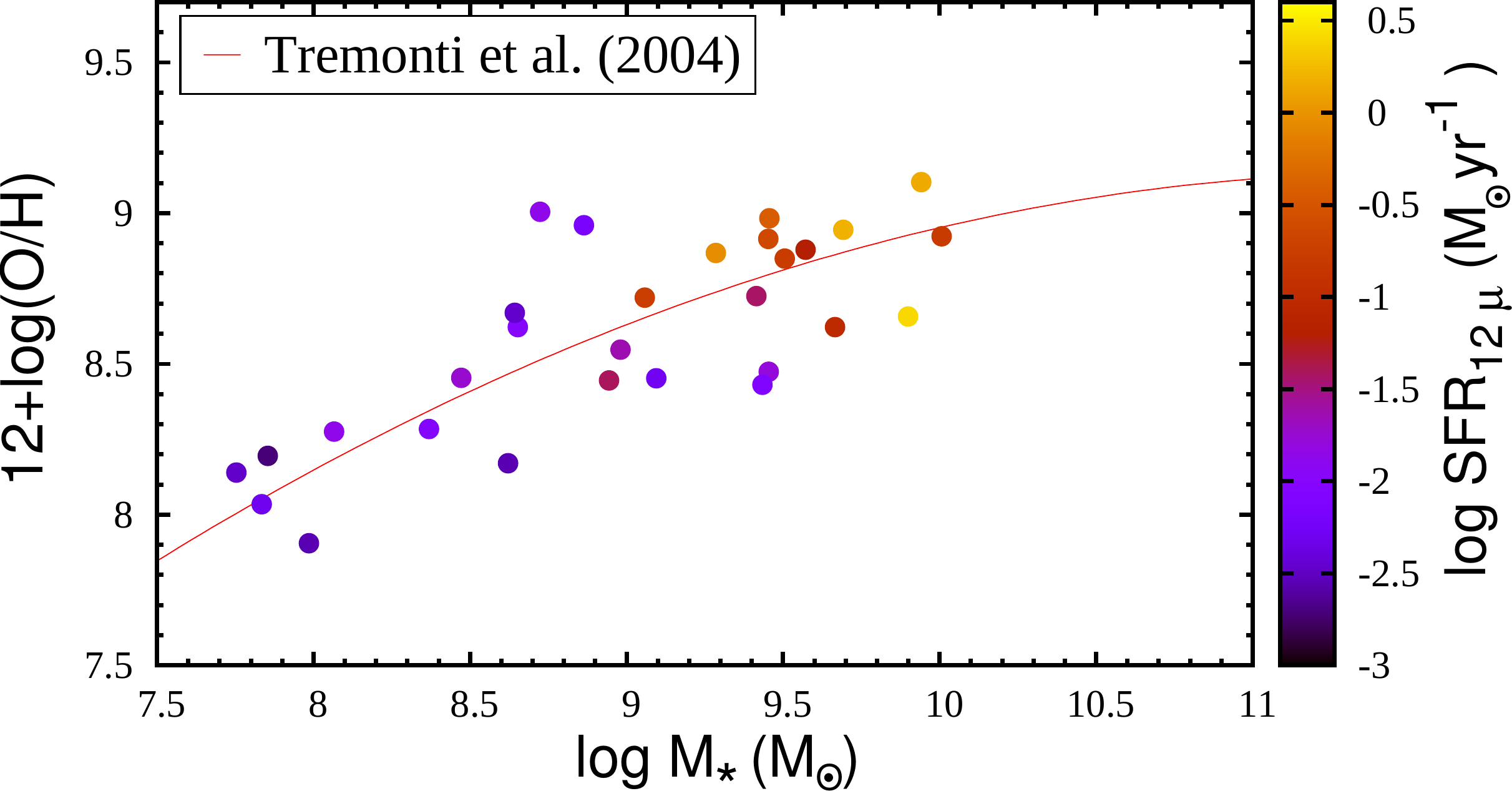} 
\label{}
}
  \caption{Mass-metallicity relation for the VGS galaxies as a function of $\rm{SFR\alpha}$ (a) and $\rm{SFR_{12}}$ (b). Mass-metallicity relation of \citet{tremonti2004}
  is indicated as a red line. }
  \label{fig:mass_metal}
\end{figure*}
\subsection{Metallicity and mid-infrared star formation rate }
\label{sec:infrared}
The Spitzer 24 $\rm{\mu}m$ band is sensitive to warm dust emission and a good tracer of star formation \citep{calzetti2007,rieke2009}. As shown in \cite{jarrett2013} there is a tight correlation 
between WISE 22 $\rm{\mu}m$ and Spitzer 24 $\rm{\mu}m$ luminosities since they 
are close in wavelength\footnote{The WISE 22 $\rm{\mu}m$ band is closer to 23 $\rm{\mu}m$ (hence, very similar to Spitzer MIPS 24 $\rm{\mu}m$), as determined by \cite{brown2014}.}. The WISE 12$\rm{\mu}m$ band, on the other hand, is more sensitive than the WISE 
22 $\rm{\mu}m$ band and probes the ISM. It is also a good indicator of star formation as it is dominated by 11.3 $\rm{\mu}m$ PAH feature that is excited by 
ultraviolet radiation from young stars, as well as radiation from older stars (\cite{jarrett2013,cluver2014} and references therein). \cite{lee2013}
also showed that both 12$\rm{\mu}m$ and 22 $\rm{\mu}m$ luminosities correlate well with $\rm{H\alpha}$ SFR estimates, concluding that both bands are 
good SFR indicators of dusty galaxies, but suffer from a metallicity bias. It is therefore important to examine metallicity as well. In this section we study the relation between the 
$\rm{H\alpha}$ and the mid-infrared star
formation rates, and the metallicity of the VGS galaxies.
\begin{figure*}
\centering
\subfigure[]
{
 \includegraphics[scale=0.61]{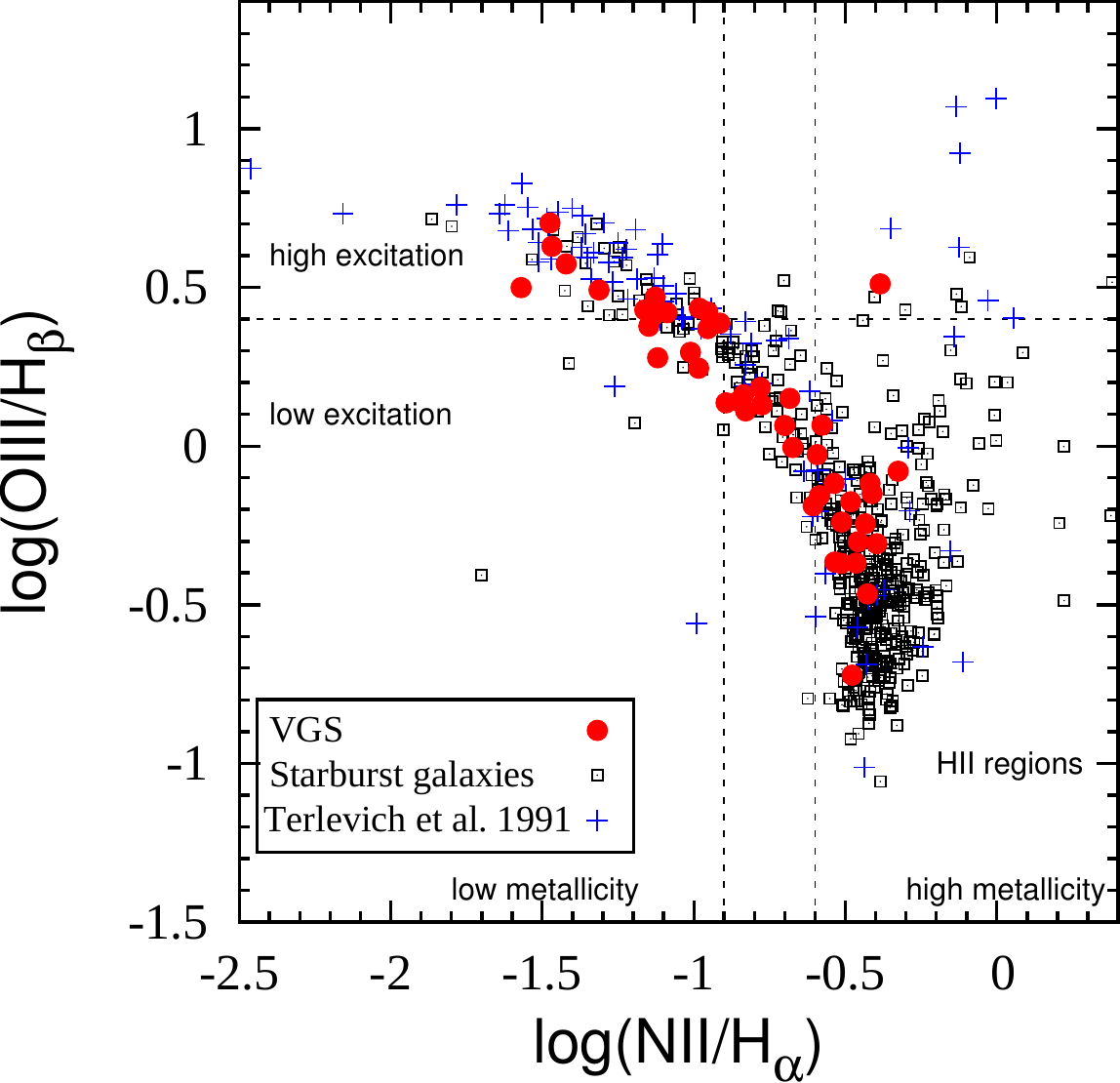} 
\label{fig:first_sub}
}
\subfigure[]
{
 \includegraphics[scale=0.42]{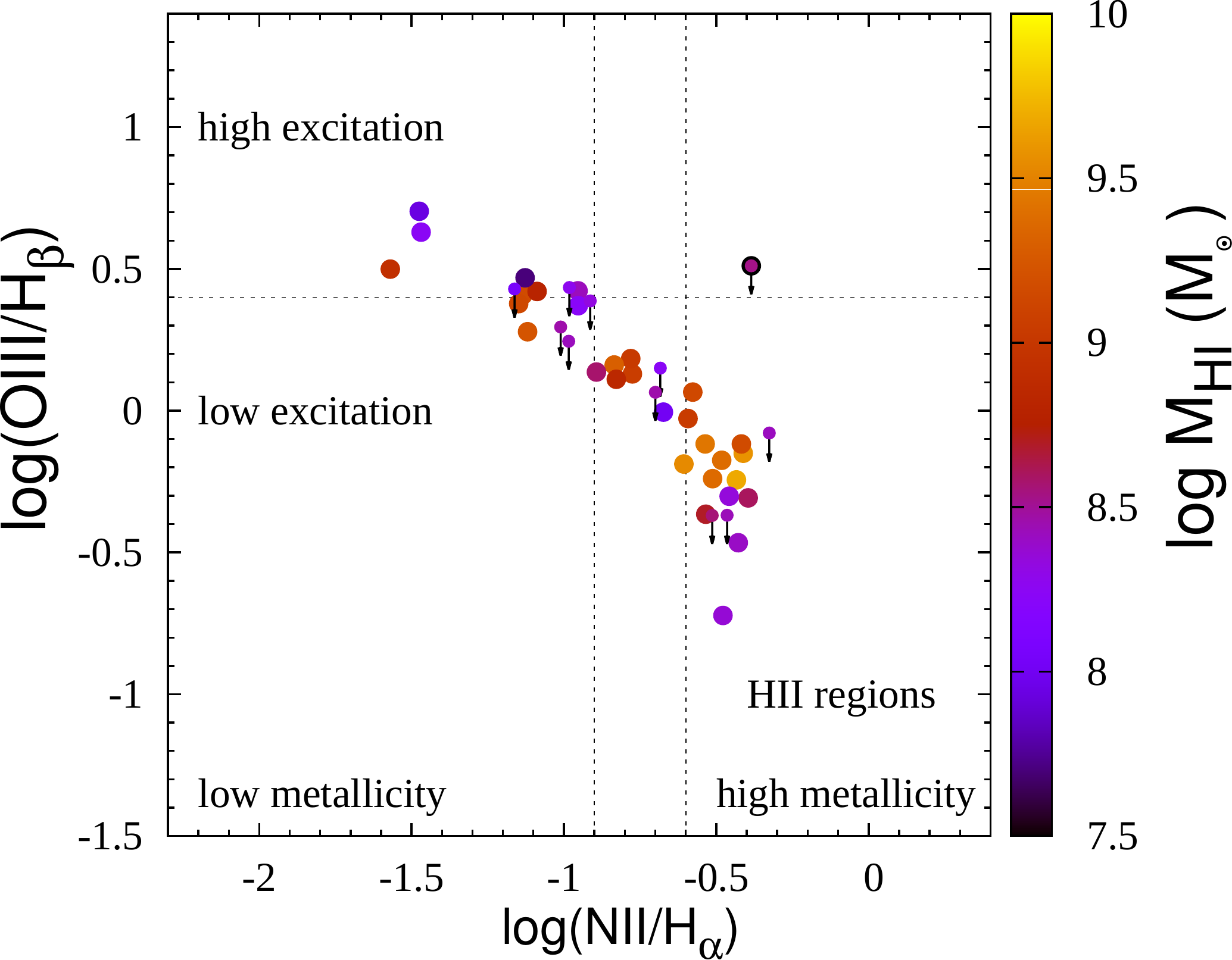} 
\label{fig:second_sub}
}
  \caption{Left: The BPT diagram showing the ratio of emission line fluxes of [O III]/$\rm{H_{\beta}}$ to [N\textsc{ii}]/$\rm{H\alpha}$ of the VGS galaxies. Red filled circles
are VGS galaxies selected as explained in section 3.3.1. The comparison sample of emission-line galaxies has been constructed from the 
emission-line galaxy sample of \citet{terlevich1991}. 
Galaxies defined as starbursts are taken from the SIMBAD and from \citet{coziol2003} where in the latter, they are defined as starburst 
Markarian galaxies. This diagram is adapted from \citet{raimann2000}. The point at the top right of 
the panel is the only AGN, VGS\_24, in our VGS galaxy sample. Right: Same as left panel, except that only VGS galaxies are shown as a function of the logarithm of their $\rm{H\textsc{i}}$
 masses. Galaxies with upper $\rm{H\textsc{i}}$ detection limit are shown with arrows. }
  \label{figure:9}
\end{figure*}
In Figure~\ref{figure:sf_w4} we examine the relation between the star formation rates derived from the WISE 12 $\rm{\mu}m$ and 22 $\rm{\mu}m$ 
luminosities and $\rm{H\alpha}$ following the studies mentioned above. First we compare the 22 $\rm{\mu}m$ SFR ($\rm{SFR_{22}}$) 
to $\rm{SFR\alpha}$ (Figure~\ref{figure:sf_w4}a) and then the 12 $\rm{\mu}m$ SFR ($\rm{SFR_{12}}$) to $\rm{SFR\alpha}$ 
(Figure~\ref{figure:sf_w4}b). In both diagrams we use the gas-phase metallicity, 12+Log(O/H), taken from the MPA/JHU catalogue for
SDSS DR7 (see $\S\ref{sec:sdss}$), as a third parameter. Galaxies that are not present in this catalogue due to low S/N on the emission lines are shown as 
black squares. Both $\rm{SFR_{22}}$ and $\rm{SFR_{12}}$ correlate well with $\rm{SFR\alpha}$ as found by \cite{lee2013} and \cite{cluver2014}. 
At low star formation rates, around $\rm{log SFR\alpha \leq -0.5}$ $\rm{M_{\odot} yr^{-1}}$, $\rm{SFR_{12}}$ tends to underestimate the SFRs. This is 
also observed by
\cite{lee2013} and \cite{cluver2014}. A possible explanation for this is the effect that low-metallicity could change the relative abundance of PAH molecules to big grains, and thereby affecting the total 
emission \citep{lee2013,cluver2014}. Indeed the VGS galaxies in the low 
$\rm{SFR_{12}}$ range have lower metallicities. On the other hand this 
discrepancy is not observed between $\rm{SFR_{22}}$ and $\rm{SFR\alpha}$ by either of these authors. 

The mass-metallicity-star formation relation of the VGS galaxies shown in Figure~\ref{fig:mass_metal} supports the observed trend in Figure~\ref{figure:sf_w4}.
Here we plot the mass-metallicity relation as a function of both $\rm{SFR\alpha}$ (Figure~\ref{fig:mass_metal}a) and $\rm{SFR_{12}}$ (Figure~\ref{fig:mass_metal}b).
In both diagrams VGS galaxies follow the mass-metallicity relation of \cite{tremonti2004}; metallicity decreases with decreasing mass. Galaxies with low star 
formation rates have lower metallicity. This trend is clearer in the 12 $\rm{\mu}$m than in the $\rm{H\alpha}$ regime where galaxies with similar $\rm{SFR\alpha}$ have
different mass. On the other hand at 12 $\rm{\mu}$m galaxies are uniformly distributed in this linear relation with respect to their stellar mass.
In conclusion, the relation between the mid-infrared and optical star formation rate tracers suggests that the VGS sample consists of both dusty high metallicity galaxies and less dusty low metallicity galaxies with in general lower star formation rates.
\begin{figure}
 \centering\includegraphics[scale=0.34]{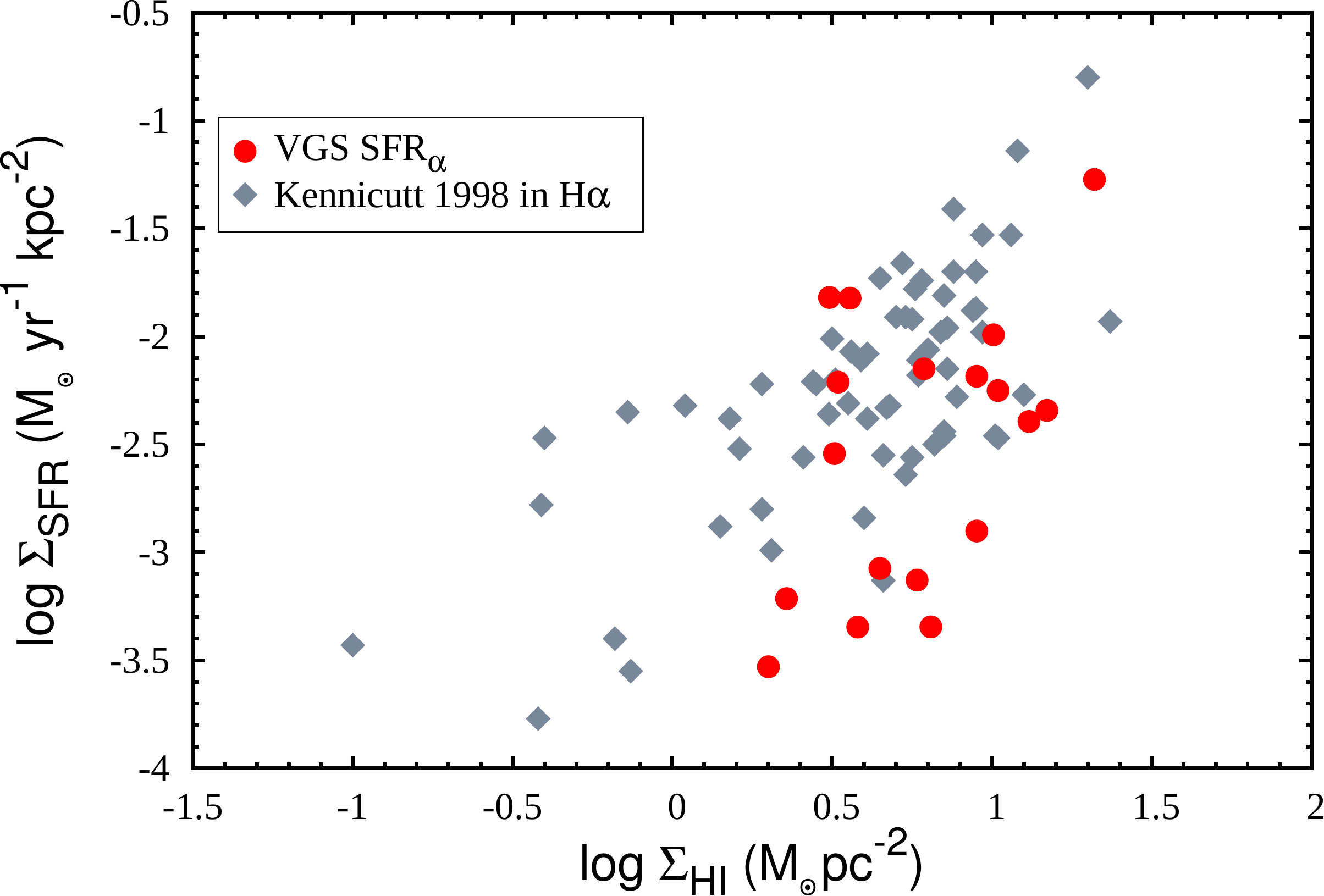} 
  \caption{Correlation of the disk-averaged $\rm{H\alpha}$ (red dots) SFRs with the surface densities of $\rm{H\textsc{i}}$. Grey diamonds indicate the spiral 
galaxies from \citet{kennicutt1998}. }
  \label{figure:KS}
\end{figure}
As for the emission line properties, VGS galaxies follow the same characteristics as other star forming galaxies, even including having an occasional AGN
(1 out of 59) and exhibiting similar proportions of normal star forming galaxies and starbursts, i.e. based on emission line
ratios when put in a BPT diagram (Figure~\ref{figure:9}a). Figure~\ref{figure:9}b shows the VGS galaxies
colour coded as a function of  their total $\rm{H\textsc{i}}$ 
mass. It is clear that the inferred excitation does not depend on  $\rm{H\textsc{i}}$ mass. 

Eleven galaxies in the low metallicity - high excitation zone have similar 
properties to the starburst dwarf galaxies. They are not detected in 22 $\rm{\mu}$, suggesting
they don't have much dust. In their spectra, they exhibit very strong $\rm{H\alpha}$ and [O III] emission lines
\footnote{Their apparent sizes are $\sim$ of 3$''$ SDSS fiber aperture, so 
the SDSS spectra can be considered as representative of the whole galaxy disk.}. Given these facts, they seem to be metal poor, and 
their position in the BPT diagram ($\rm{log(NII/H\alpha)} < -0.9$) confirms this (Figure~\ref{figure:9}a). 

\subsection{Dependence of star formation on $\rm{H\textsc{I}}$ surface density}

The Kennicutt-Schmidt (KS) law \citep{kennicutt1998} describes the relation between the surface density of star formation and the surface density of gas
 ($\rm{H\textsc{i}}$ + $\rm{H_{2}}$). The KS law was originally applied to global properties, but has also been demonstrated (\cite{bigiel2008} and \cite{kenev2012}) to describe the local relation between SF surface 
density and gas density within galaxies. Since we do not know the
 $\rm{H_{2}}$ content of our galaxies (except VGS\_31a \citep{beygu2013}) and our $\rm{H\textsc{i}}$ data is not good enough for a spatial KS law analysis we examined the global 
KS law for the VGS galaxies. To do this we have selected 18 VGS galaxies within our sample for which we can resolve the $\rm{H\textsc{i}}$ surface density distribution within the galaxy 
disk\footnote{\cite{kreckel2012} found that most of the VGS galaxies have $\rm{H\textsc{i}}$ disks two to three times more extended than their stellar disks.}. 
We calculated the average surface densities of star formation and $\rm{H\textsc{i}}$ within the Petrosian R90 radius taken from SDSS DR7 (the radius containing
 90 $\% $ of Petrosian flux \citep{petrosian1976}). The Petrosian R90 radius can be considered a representative radius for calculating average surface densities since the $\rm{H\alpha}$ emission 
of the VGS galaxies does not extend beyond this radius. In Figure~\ref{figure:KS} we show the global Kennicutt-Schmidt law for the 18 resolved VGS galaxies 
determined from the $\rm{H\alpha}$ and also compare them with the sample of spiral galaxies from \cite{kennicutt1998}. 

Figure~\ref{figure:KS} covers the range of $\rm{\Sigma_{SFR}}$ and $\rm{\Sigma_{H\textsc{i}}}$ where normal and low surface brightness galaxies are mostly located.  
We do not see a strong correlation
between the $\rm{\Sigma_{SFR}}$ and the $\rm{\Sigma_{H\textsc{i}}}$ of the VGS galaxies but rather a large spread. A similar behaviour is shown by the low surface brightness
galaxies in \cite{kennicutt1998}. For the spiral galaxy sample studied in \cite{kennicutt1998}, even after including the $\rm{H_{2}}$ to the gas density, a large scatter is still
present (Figure 2 in that paper). In the same paper an attempt was made to examine the SFR versus the $\rm{H\textsc{i}}$ density relation for a larger sample of spiral 
galaxies (Figure 4 in that paper), however the large dispersion is still present. The correlation is very similar to  Figure~\ref{figure:KS} as well and
as \cite{kennicutt1998} stated the physical interpretation of the SFR versus $\rm{H\textsc{i}}$ Schmidt law is not obvious, and complicated by the mutual affects of the UV field associated 
with the recent star formation and the balance between the atomic and molecular fraction of the neutral hydrogen in this low density regime.

\begin{figure}
 \centering\includegraphics[scale=0.44]{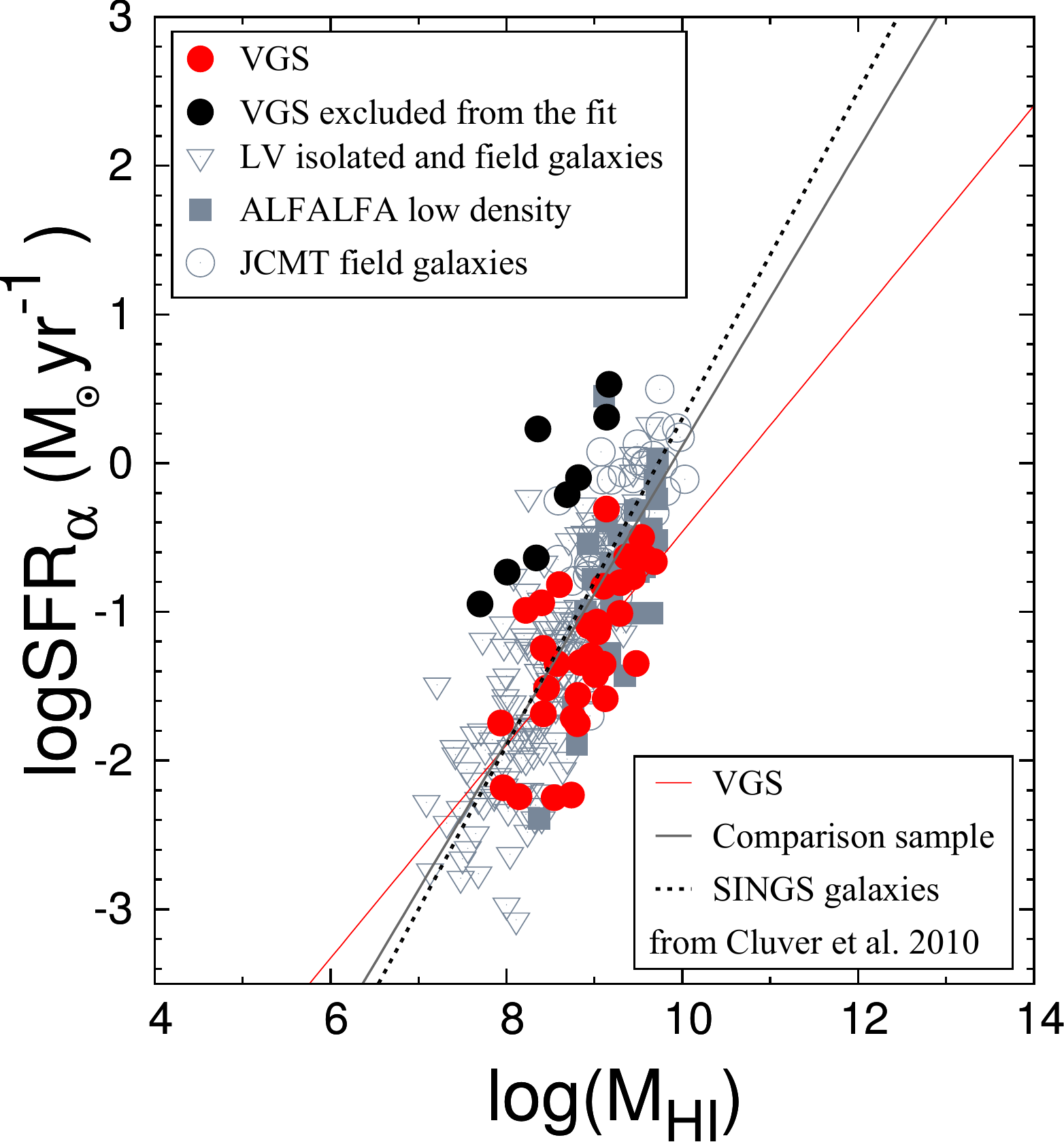} 
  \caption{$\rm{SFR_{\alpha}}$ against the total $\rm{H\textsc{i}}$ mass of the VGS galaxies, the comparison sample and the SINGS galaxies from \citet{cluver2010}. Linear least-squares fits are made through the VGS galaxies (red solid line), the comparison sample (grey solid line) and the SINGS galaxies (black dotted line). VGS galaxies that have significantly elevated SFR for a given $\rm{H\textsc{i}}$ mass (black dots) are excluded from the fit.}
  \label{figure:sfr_tot_gas}
\end{figure}

\begin{figure*}
\centering
\subfigure[]
{
  \centering\includegraphics[scale = 0.33]{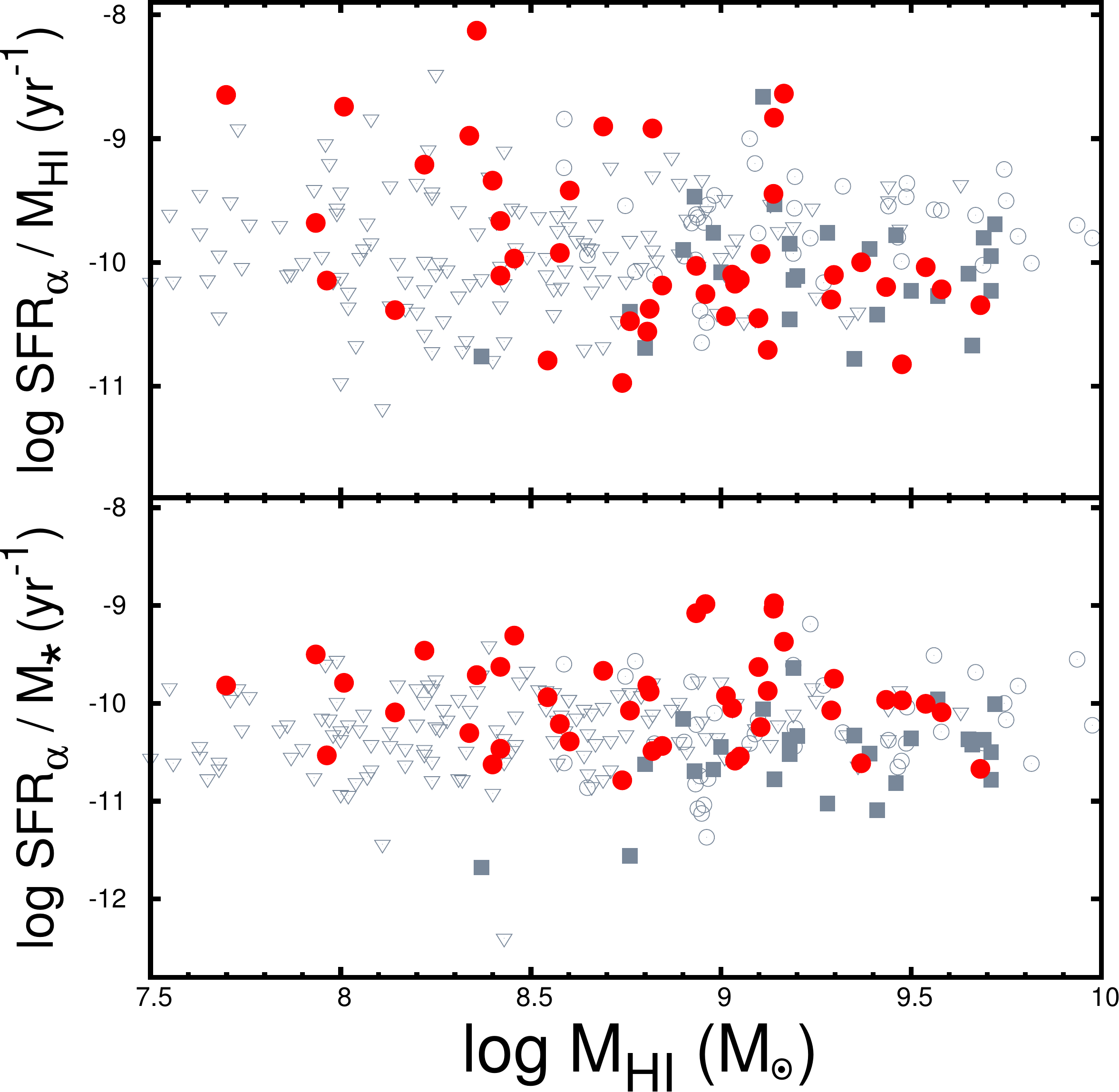} 
}
\subfigure[]
{
  \centering\includegraphics[scale = 0.33]{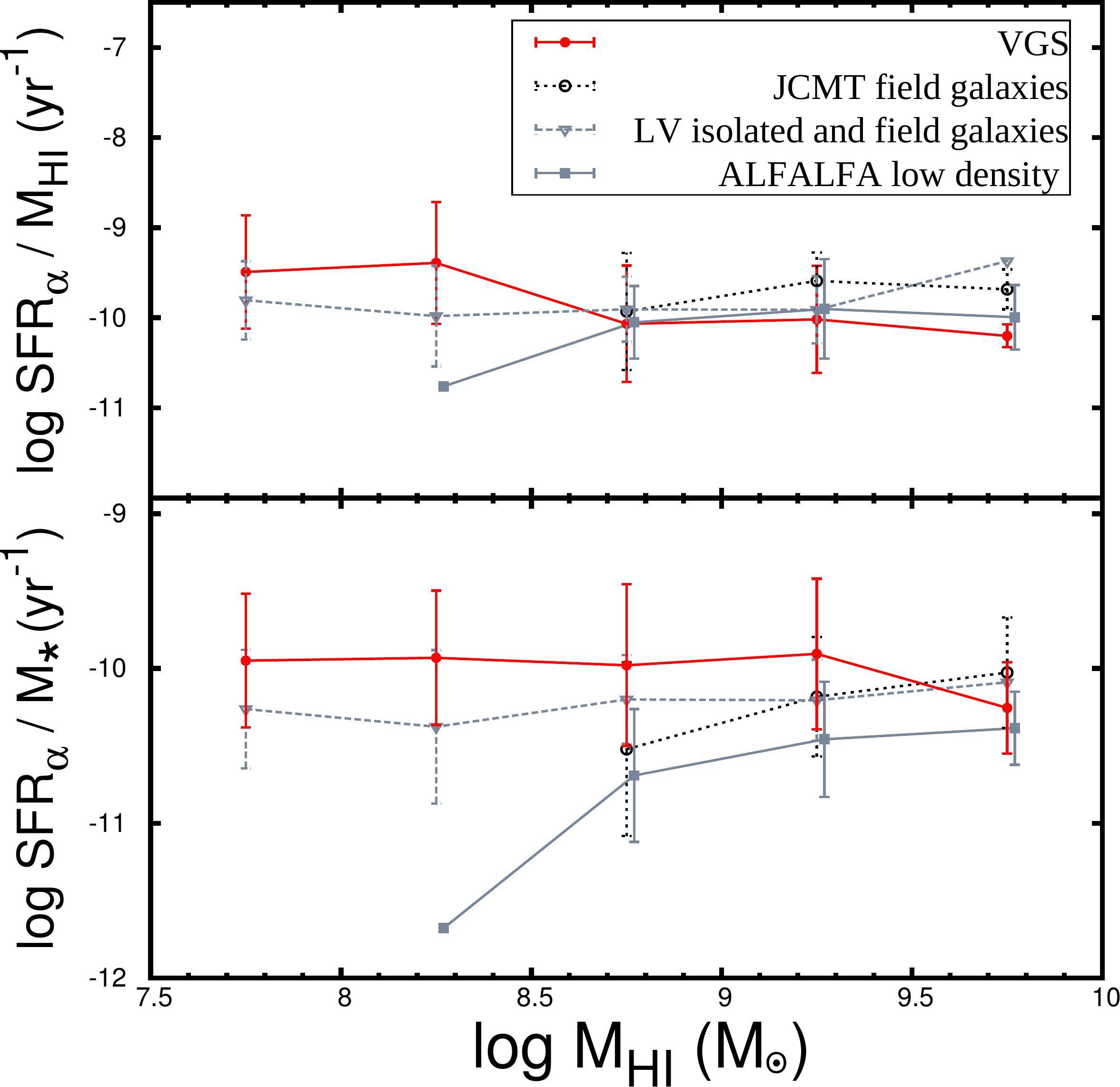} 
}
  \caption{ Left panel: The extinction corrected $\rm{SFR\alpha}$ normalized by the $\rm{M_{HI}}$ (top) and normalised by the $\rm{M_{*}}$ (bottom) are plotted against the $\rm{M_{HI}}$ for the VGS galaxies
and the comparison sample. Right panel: The mean of the  $\rm{SFR\alpha}$ / $\rm{M_{HI}}$ and  $\rm{SFR\alpha}$ / $\rm{M_{*}}$ are plotted per 0.5 dex of $\rm{M_{HI}}$ (only for 
detections) for the distributions shown on the left panel. }
 \label{figure:11}
\end{figure*}

In Figure~\ref{figure:sfr_tot_gas} we show the relation between the $\rm{H\alpha}$ star formation rate and the total $\rm{H\textsc{i}}$ mass for the VGS galaxies and the comparison sample. We also plot additional data
from The Spitzer Infrared Nearby Galaxies Survey (SINGS) galaxy sample \citep{cluver2010} that consists of infrared imaging and spectroscopic survey of 75 nearby galaxies. In Figure~\ref{figure:sfr_tot_gas} three 
linear fits are shown; the VGS galaxies (red solid line), the comparison sample together (grey solid line) and the SINGS galaxies (black dotted line). We excluded 8 VGS galaxies (shown as black dots) 
with $\rm{Log SFR\alpha/M_{HI}}$ $>$ $\rm{10^{-9} yr^{-1}}$ from the fit. These are the objects that have significantly elevated SFR for a given $\rm{H\textsc{i}}$ mass, also clearly seen in the top panel 
of Figure~\ref{figure:11}a. Overall each comparison sample display a linear relation similar to SINGS galaxies \citep{cluver2010}. VGS galaxies, however, scatter above the relation indicating more elevated SFR for a given total $\rm{H\textsc{i}}$ mass. This indicates that the VGS galaxies have higher star formation rate for their total $\rm{H\textsc{i}}$ content. We will discuss this further in the next section.


 \begin{figure*}
\centering
\subfigure[]
{
  \centering\includegraphics[scale = 0.32]{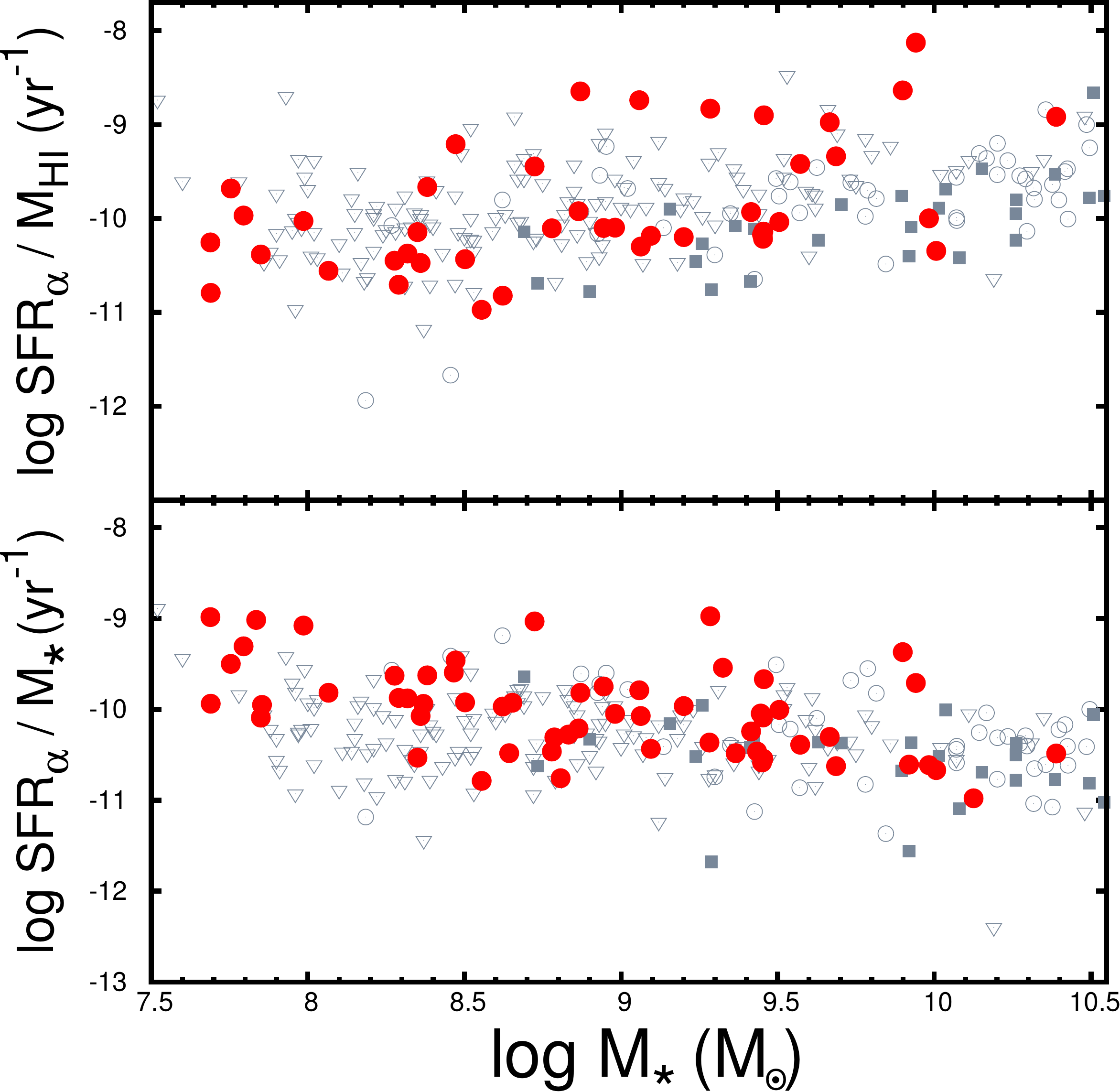} 
}
\subfigure[]
{
  \centering\includegraphics[scale = 0.319]{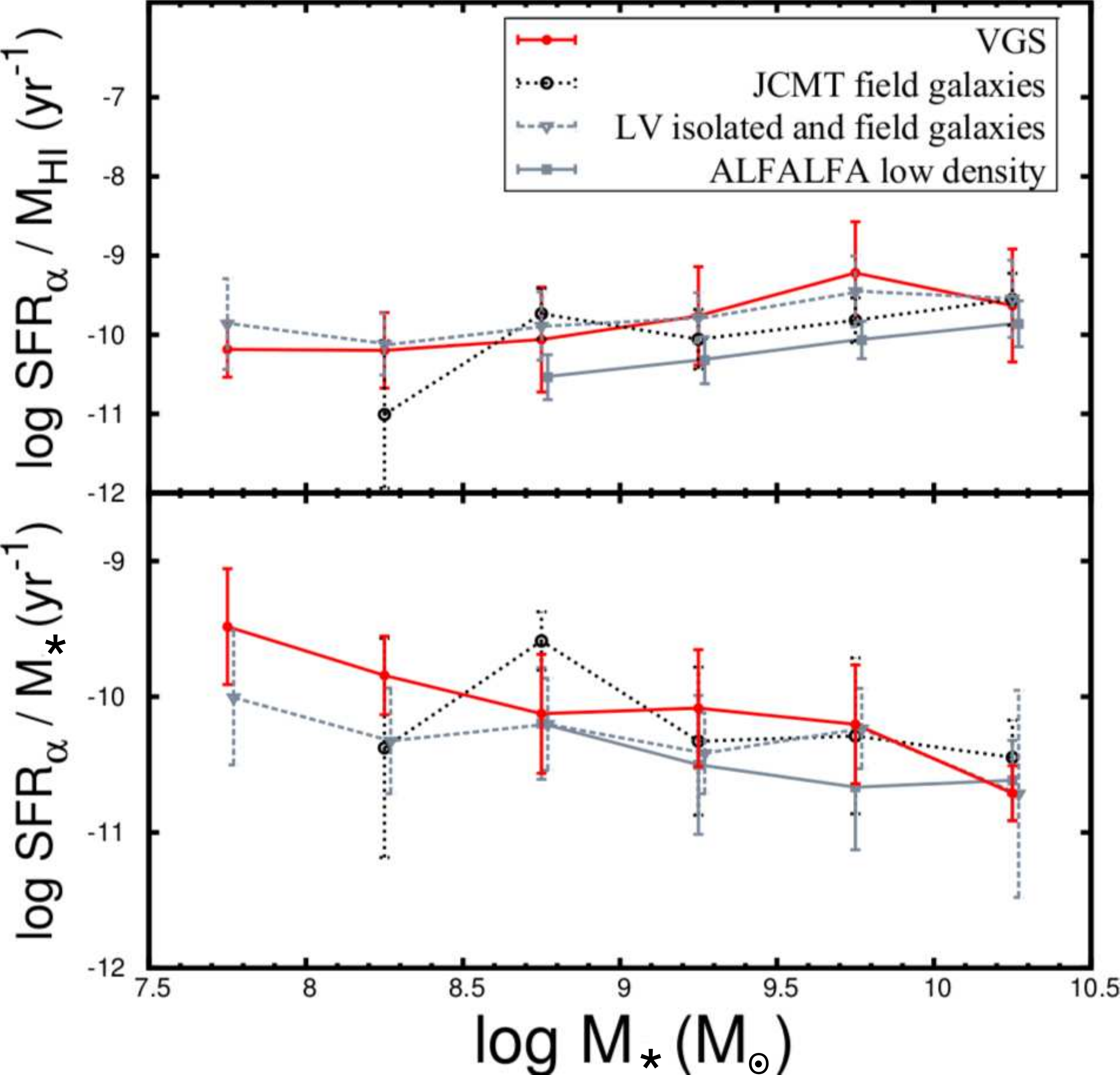}
}
\caption{ Left panel: The extinction corrected $\rm{SFR\alpha}$ normalised by the $\rm{M_{HI}}$ (top) and normalised by the $\rm{M_{*}}$ (bottom) are plotted against the $\rm{M_{*}}$ for the VGS galaxies
and the comparison sample. Right panel: The mean of the  $\rm{SFR\alpha}$ / $\rm{M_{HI}}$ and  $\rm{SFR\alpha}$ / $\rm{M_{*}}$ are plotted per 0.5 dex of $\rm{M_{*}}$ (in the 
$\rm{SFR\alpha}$ / $\rm{M_{HI}}$ case, only for the $\rm{H\textsc{i}}$ detections) for the distributions shown on the left panel. }
 \label{figure:12}
\end{figure*}

\subsection{Specific star formation and star formation efficiencies.}

For a proper comparison of star formation properties it is important to
normalise the observed star formation rates. This can be done in different 
ways. It has become customary to normalise to the stellar mass (giving a
specific star formation rate) or to the gas mass (giving a star formation
efficiency). Here we use both measures for both the VGS galaxies and the 
galaxies in the comparison samples, making sure that the star formation rates
and stellar masses have been determined using the same methods.

In Figures~\ref{figure:11} and~\ref{figure:12} we compare the 
specific star formation rate ($\rm{SFR\alpha/M_{*}}$) and star formation efficiency ($\rm{SFR\alpha/M_{HI}}$) of the VGS galaxies to those of field galaxies from the comparison samples. 

It is clear from Figures~\ref{figure:11} and~\ref{figure:12} that
$\rm{SFR\alpha/M_{*}}$ and $\rm{SFR\alpha/M_{HI}}$ of the VGS galaxies 
as a function of $\rm{M_{*}}$ and 
$\rm{M_{HI}}$ behave in a similar way as those of the field galaxies 
in our comparison samples. In 
Figures~\ref{figure:11}b and~\ref{figure:12}b, we plot the mean of the $\rm{SFR\alpha/M_{*}}$ and 
$\rm{SFR\alpha/M_{HI}}$ per 0.5 dex of  $\rm{M_{HI}}$ and 
$\rm{M_{*}}$. Although not each mass bin is uniformly filled both by the VGS and the comparison sample galaxies, it is still possible to 
derive statistical information. Below we discuss $\rm{SFR\alpha/M_{HI}}$ and $\rm{SFR\alpha/M_{*}}$.

In Figure~\ref{figure:11}, the VGS and the comparison sample galaxies have similar $\rm{SFR\alpha/M_{HI}}$ and $\rm{SFR\alpha/M_{*}}$ as a function of $\rm{M_{HI}}$, except at $\rm{M_{HI}}$ $<$ $\rm{10^{9}}$ $\rm{M_{\odot}}$ and $\rm{M_{HI}}$ $\geq$ $\rm{10^{9}}$ $\rm{M_{\odot}}$. At $\rm{M_{HI}}$ $<$ $\rm{10^{9}}$ $\rm{M_{\odot}}$, VGS galaxies have slightly higher
 $\rm{SFR\alpha/M_{HI}}$  than the comparison sample of LV galaxies which is the only sample of comparison galaxies populating this mass range. This is expected as shown in Figure~\ref{figure:sfr_tot_gas}. At 
$\rm{M_{HI}}$ $>$ $\rm{10^{9}}$ $\rm{M_{\odot}}$, VGS galaxies do not have notably different $\rm{SFR\alpha/M_{HI}}$  than the comparison sample. This mass range is mostly populated by 
the JCMT galaxies and subsequently by the ALFALFA galaxies. The $\rm{SFR\alpha/M_{*}}$ of the VGS galaxies appears to be slightly higher than that of the comparison samples over the entire stellar mass 
range, though the difference is very small.

When corrected for extinction  both the $\rm{SFR\alpha/M_{HI}}$ and $\rm{SFR\alpha/M_{*}}$ of the VGS galaxies and the comparison sample galaxies as a function 
of $\rm{M_{*}}$ (Figure~\ref{figure:12}) are very similar, except that 
 two VGS galaxies around $\rm{M_{*}} = \rm{10^{10}}$ $\rm{M_{\odot}}$ stand out with higher $\rm{SFR\alpha/M_{HI}}$ (Figure~\ref{figure:12}a, top panel). These are VGS\_31a and VGS\_57.

Overall, the star formation efficiency is very constant over the range of stellar masses covered while the specific star formation rate decreases with increasing stellar mass, as already noted by 
\cite{kreckel2012} using SFRs extrapolated from the $\rm{H\alpha}$ fluxes from the SDSS DR7 fiber spectra. The decrease in $\rm{SFR\alpha/M_{*}}$ is about a factor ten over a hundredfold increase in 
stellar mass from $\rm{10^{8}}$ to $\rm{10^{10}}$ $\rm{M_{\odot}}$. This trend is not specific to galaxies in voids but also present in galaxies in slightly denser environments. 

\section{Discussion}

Voids and void galaxies are very suitable candidates to test and study two of the most important theories on galaxy formation and evolution.
These are related to the formation of dwarf galaxies and the role of cold gas accretion.

Although void galaxies in our sample are mostly blue and small galaxies with stellar masses below 3 $\rm{\times}$ $\rm{10^{10}}$ $\rm{M_{\odot}}$, 
we do not detect the missing dwarf galaxy population \citep{kreckel2012} whose absence was stressed  by \cite{peebles2001} as a major riddle for our understanding of galaxy formation.
A simple observational ground for this could be the difficulty in detecting these objects at this redshift range 
(0.02 $<$ z $<$ 0.03). The reason why these galaxies are not included in our VGS sample is that they remain undetected as the VGS galaxies have been selected within the SDSS
 spectroscopic flux limit of 17.7 mag in r band.
On the other hand, if their absence in the VGS sample is the reflection of a true physical phenomenon, it may be the result of some interesting astrophysical process. One of the most 
suggestive one is the proposition of \cite{hoeft2010}. They argue that photo-heating due to the cosmic UV background stops gas condensation in the halos of these small galaxies.
In their simulations, \cite{hoeft2010} show that the already condensed gas can be retained by these halos. This gas would be allowed further star formation. However, as mentioned 
above, our observations may simply not be sensitive enough to detect these dwarf galaxies affected by photo-heating.

Another interesting physical process was first discussed by \cite{dekel2006}. They proposed that the galaxy bimodality (blue and red sequence ) in the colour-magnitude space could be 
understood by looking at the thermal properties of the inflowing gas, in particular the 
role and presence of cold accretion flows. According to their model, the blue sequence galaxies below a characteristic stellar mass 
$\rm{M_{*,crit}}$ $\simeq$ 3 $\rm{\times}$ $\rm{10^{10}}$ $\rm{M_{\odot}}$, reside in halos whose mass is below a critical shock-heating mass. In these halos, the disc is built by cold
 flows and might generate early starbursts \citep{dekel2006}. 

Simulations by \cite{keres2005} also suggested that low density regions today are dominated by cold mode accretion. Since void galaxies in our 
sample are all below $\rm{M_{*,crit}}$, this would suggest that cold accretion dominates. It would imply that they to form an ideal sample to study cold
 flow dominated galaxy evolution further. Moreover, the tenuous large-scale structure around void galaxies determines the way a galaxy may accrete mass in terms of direction and more 
importantly, the coherence of the infalling matter \citep{aragon2013,rieder2013}.

Our study indicates that galaxies in voids are similar in their star formation properties to galaxies in 
slightly denser environments in the nearby universe. The only notable difference is that the specific star formation rate maybe slightly enhanced by $~$0.2 dex. This may indicate a slightly more steady (and less disrupted) inflow of gas in the picture of cold accretion. 
To put the star formation properties of void galaxies in the context of their formation history, we plan to study in more detail their molecular hydrogen content as well as their metallicity
and their star formation histories. This will allow us to put further constraints on the place of these void galaxies in the general context of galaxy formation. In an accompanying paper we will
discuss the morphology and colour properties of these void galaxies (Beygu et al. to be submitted)

\section{Conclusions}

We have examined the star formation properties of the VGS galaxies and compared them with field galaxies. The main conclusions are:

\begin{itemize}
 
 \item The specific star formation rate of the VGS galaxies follows the same decreasing trend with stellar mass as galaxies in somewhat denser environments 
 (Figure~\ref{figure:11} and~\ref{figure:12}). The star formation efficiencies are fairly constant with stellar mass, although their SFR is slightly elevated at $\rm{M_{HI}}$ $<$ $\rm{10^{9}}$ $\rm{M_{\odot}}$ for their total $\rm{H\textsc{i}}$ content. These are more difficult to interpret as 
 explained in section 3.6.4. The specific star formation rate with stellar mass $\rm{M_{\odot}}$, is slightly elevated in the void galaxies.

\item VGS galaxies display a wide range in dust content and metallicity. The sample consists of dusty and high metallicity galaxies as well as galaxies
with insignificant dust content and low metallicity.

 \item The VGS galaxies appear to obey the canonical 'star formation main sequence' uncovered in galaxies in general \citep{kennicutt1998,tremonti2004}, indicating that the star formation process progresses
 under similar physical conditions as in galaxies in environments somewhat denser than voids. 

 \item There is no strong correlation between the $\rm{H\alpha}$ formation properties of the VGS galaxies and 
 their $\rm{H\textsc{i}}$ content.

\end{itemize}

\section*{Acknowledgements}

The authors wish to thank Michael Vogeley, for a careful assessment of the manuscript and very useful comments. BB is grateful to Jarle Brinchmann and Daniela Calzetti
for many helpful discussions and insightful thoughts and also wishes to thank I.D. Karachentsev and J. R. S\'{a}nchez-Gallego for their help in 
use of their catalogues, and to thank Marius Cautun, Peppo Gavazzi and Manolis Papastergis. We would like to thank M. Querejeta and S. Meidt for sharing their S4G nearby galaxy stellar mass catalogue.
This work was supported in part by the National Science Foundation under grant no. 1009476 to Columbia University. We are grateful 
for support from a Da Vinci Professorship at the Kapteyn Astronomical Institute. 
J.M. van der Hulst acknowledges support from the European Research Council under the European Union's Seventh Framework Programme (FP/2007-2013)/ ERC Grant Agreement nr. 291531. K. Kreckel acknowledges 
grants 
4598/1-2 and SCHI 536/8-2 from the DFG Priority Program 1573. MDM observatory is located on the
southwest ridge of Kitt Peak, home of the Kitt Peak National Observatory, Tucson, Arizona. The Observatory is owned and operated 
by a consortium of five universities: the University of Michigan, Dartmouth College, the Ohio State University, Columbia University, 
and Ohio University. We acknowledge KPNO for the use of their $\rm{H\alpha}$ filters.
The Isaac Newton Telescope is operated on the island of La Palma by the Isaac Newton Group 
in the Spanish Observatorio del Roque de los Muchachos of the Instituto de Astrof\'{\i}sica de Canarias.



\bibliographystyle{mnras}
\bibliography{beygu} 

\bsp	
\label{lastpage}
\end{document}